\documentclass[aps,showpacs,twocolumn,superscriptaddress]{revtex4}
\usepackage{epsfig}
\usepackage{graphics}
\usepackage{amssymb}
\usepackage{amsmath}

\newcommand{\intsum}[1]{\sum_{#1} \! \! \! \! \! \! \! \! \! \int }
\begin{document}
\title{Many-body calculations of two-photon, two color matrix elements for attosecond delays}

\author{Jimmy \surname{Vinbladh}}
\affiliation{Department of Physics, Stockholm University, AlbaNova University Center, SE-106 91 Stockholm, Sweden}
\author{Jan~Marcus \surname{Dahlstr\"om}}
\affiliation{Department of Physics, Lund University, Box 118, SE-221 00 Lund, Sweden}
\author{Eva \surname{Lindroth}}
\email{Eva.Lindroth@fysik.su.se}
\affiliation{Department of Physics, Stockholm University, AlbaNova University Center, SE-106 91 Stockholm, Sweden}

\begin{abstract}
We present calculations for attosecond  atomic delays in photoionization of noble gas atoms based on full two-color two-photon Random-Phase Approximation with Exchange  in both length and velocity gauge. Gauge invariant atomic delays are demonstrated for the complete set of diagrams.   The results are used to investigate the  validity of the common assumption that the measured atomic delays can be interpreted as a one-photon Wigner delay and a universal continuum--continuum contribution that depends only on the kinetic energy of the photoelectron, the laser frequency  and the charge of the remaining ion, but not on the specific atom or the orbital from which the electron is ionized. Here we find 
that although effects beyond the universal IR--photoelectron continuum--continuum transitions are rare, they do occur in special cases such as around the $3s$ Cooper minimum in argon.
We conclude also that in general the convergence in terms of many-body diagrams is considerably faster in length gauge than in velocity gauge.
\end{abstract}

\maketitle


\section{Introduction}
Techniques for probing ultrafast electronic dynamics, such as  the Reconstruction of Attosecond  Beating By Interference of Two-photon Transitions (RABBIT)~\cite{PaulScience2001} or the attosecond streak camera~\cite{ItataniPRL2002},
use delay-dependent modulations in photoelectron spectra  to quantify the time it takes for an electron to escape an atomic potential~
\cite{SchultzeScience2010,KlunderPRL2011,GuenotPRA2012,GuenotJPB2014,Sabbar:2015,Kotur:2016,GrusonScience2016,Isinger2017,CirelliNatCommun2018}. 
These modulations arise since the interaction with the ionizing attosecond pulse (or pulse train) takes place in the presence of a laser field that is phase-locked to the attosecond light field(s). 
It has further  been established~\cite{DahlstromJPB2012,DahlstromCP2013,PazourekFD2013} that it is meaningful 
 to separate the measured atomic delay, $\tau_A\approx\tau_{W}+\tau_{cc}$, into a Wigner-like delay associated with the one-photon extreme ultraviolet (XUV) ionization process, $\tau_W$, and a contribution from the interaction with the infrared (IR) laser-field in the presence of the atomic potential, called the continuum--continuum delay, $\tau_{cc}$ (or Coulomb-Laser Coupling delay in the context of streaking). In this context $\tau_{cc}$ denotes the contribution from a single photoelectron in a Coulomb field that absorbs or emits an IR photon, as detailed in ~\cite{DahlstromJPB2012,DahlstromCP2013}.  While the Wigner delay is known to be strongly dependent on the atomic origin of the electron \cite{KheifetsPRA2013}, the contribution from the IR photon has been found to be more ``universal'' in the sense that it depends only on the kinetic energy of the photoelectron, the photon energy of the laser field and the charge of the remaining ion \cite{DahlstromCP2013}. 
In the limit of weak fields, the physics can be described as interference effects between various two-photon processes.  
The validity of the $\tau_{cc}$ correction has been studied through  many-body calculations of the two-photon process both for angular integrated measurements and for detection along the polarization axis~\cite{DahlstromPRA2012,DahlstromJPB2014}. 
Although there are exceptions, in particular close to ionization thresholds and at resonances~\cite{DahlstromJPB2014,Sabbar:2015,Kotur:2016}, the universality of the contribution from the laser photon has hitherto proved to be a good approximation, with the important practical consequence that Wigner delays, $\tau_W$, can be extracted from measured atomic delays, $\tau_A$. The many-body calculations themselves have been benchmarked, for example against the  difference between $2s$ and $2p$ time delays in neon~\cite{Isinger2017}, over a wide energy range, and against time delay differences between the outermost shells of  rare gas atoms~\cite{GuenotJPB2014}.
Still, 
the procedure used so far has  employed significant  approximations. 
First, only the dominating time-order with the XUV photon being absorbed first, and the IR photon being exchanged subsequently in a continuum-continuum transition, has been studied in much detail \cite{LindrothPRA2017}. 
Second, a more careful account for many-body effects has only been done for the XUV photoionization process, while the interaction with the second photon has been calculated in the lowest-order/classical approximation \cite{DahlstromPRA2012,FeistPRA2014}. 
Although these approximations are reasonable they have important consequences: the results at this level of theory are expected to depend on whether the light--matter interaction is expressed in the length or velocity gauge. In addition, experimental results on the difference between $3s$ and $3p$ time delays in argon~\cite{KlunderPRL2011,GuenotPRA2012} show a marked disagreement with theory in the region around the $3s$--Cooper minimum (at photon energies of $\sim 40$~eV). 
Therefore, it is important to push the study of atomic delays one step further. Here we have performed 
Random-Phase Approximation with Exchange (RPAE) type calculations for the {\it complete} two-photon process. The RPAE Approximation, which is identical to Time-dependent Hartree-Fock (TDHF), is known to account for the dominating many-body effects in one-photon ionization \cite{Amusia1990}. While the length and velocity form of the electric dipole interaction gives the same result for  electrons in any local potential, the use of the Hartree-Fock exchange potential destroys this invariance. As was shown more than forty years ago~\cite{lin:77}, RPAE, which accounts fully  for hole-particle excitations (including the effects usually called ground-state correlation, see below) is able to restore 
the gauge invariance. 
For two-photon processes~\cite{PhysRevA.33.3938,PhysRevA.41.6271} and beyond~\cite{sekinoandbartlett:92} pioneering studies of  many-body effects on the RPAE-level were done already in the nineteen eighties and nineties. The target at the time was  absorption of equal energy photons for below-threshold ionization (one photon alone could not induce ionization). 
In contrast, our interest is in the interaction with two photons of very different energies, where one photon can initiate an above-threshold ionization process. We will demonstrate that, just as for one-photon ionization, gauge invariance is obtained when hole-particle excitations are fully accounted for including all time orders, i.e. in a complete two-photon RPAE calculation. We further show that the size of individual contributions is vastly different in the two gauges and that the common approximation to neglect the time-order where the IR photon is absorbed first leads to wrong results in velocity gauge.

In Sec.~\ref{theory} we revisit the theory for atomic delays. The method of calculation is outlined in Sec.~\ref{method} and  the results are presented in Sec.~\ref{results}. In Sec.~\ref{sec:discussion} we present a discussion of our findings and in Sec.~\ref{sec:conclusion} we present our conclusions. 

\section{Theory}
\label{theory}
Here we will briefly discuss the calculation of delays in laser-assisted photoionization. A more detailed account can be found in Ref.~\cite{DahlstromJPB2014}. 
We consider first an $N$-electron atom that absorbs one photon and subsequently ejects a photoelectron. 
The radial photoelectron wave function will asymptotically be described by an outgoing phase-shifted Coulomb wave

\begin{align}
\label{onephoton2}
\nonumber \\
u^{(1)}_{q , \Omega, a}
\left( r \right) \approx - \pi M^{(1)} \left( q , \Omega, a \right)  \sqrt{\frac{2m}{\pi k \hbar^2}} \nonumber \\
 \times e^{i \left( k r + \frac{Z}{k a_0} \ln 2k r- \ell \frac{\pi}{2} + \sigma_{Z,k,\ell} + \delta_{k,\ell}
 \right) },
\end{align}
where $M^{(1)}$ is the electric dipole transition matrix element to the final continuum state $q$ with momenta $k$, $\ell$, and $m_a$.  When  correlation effects are accounted for $M^{(1)}$ can contribute to the phase shift, and $\sigma_{Z,k,\ell}$ is the Coulomb phase for a photoelectron with wave number $k$ and angular momentum quantum number $\ell$ in the field from a charge of  $Ze$: 
\begin{align}
\sigma_{Z,k,\ell}= \arg \left[ \Gamma  \left(  \ell + 1  -i\frac{Z}{k a_0} \right) \right].
\end{align}  
The  phase $\delta_{k,\ell}$ in Eq.~(\ref{onephoton2}) denotes the additional shift induced by the atomic potential at short range. 
 In the following we  label the full perturbed wave function associated with absorption of one photon with angular frequency $\Omega$ and a hole  in orbital $a$, by $\left|\rho_{\Omega,a}\right>$, including both radial, angular and spin parts implicitly. 

We will consider measurements that employ the RABBIT technique \cite{PaulScience2001}, where an XUV comb of odd-order harmonics of a fundamental laser field, $\Omega=(2n+1)\omega$, is combined with a synchronized, weak laser field with angular frequency $\omega$. In RABBIT, the one-photon ionization process is assisted by an IR photon that is either absorbed or emitted. This gives rise to quantum beating of sidebands in the photoelectron spectrum at energies corresponding to the absorption of an even number of IR photons.  
The outgoing radial wave function after interaction with two photons (one XUV photon, $\Omega$, and one laser photon, $\omega$), will asymptotically have the form 
\begin{align}
\label{twophoton}
u_{q, \omega,\Omega, a}^{(2)}
\left( r \right) 
\approx  
 -\pi
M^{(2)}(q, \omega, \Omega, a)\sqrt{\frac{2m}{\pi k \hbar^2}} \nonumber \\
e^{i \left( k r + \frac{Z}{k a_0} \ln kr - {\ell} \frac{\pi}{2} + \sigma_{{\ell},Z}\left( k \right) + \delta_{{\ell}}\left( k \right)
 \right) }, 
\end{align}
where the important difference compared to the one-photon case lies in the presence of the two-photon transition element $M^{(2)}$, which connects the initial state $\mid a\rangle$ to the  continuum state $\mid q \rangle$ through all dipole-allowed intermediate states. 

\subsection{The form of the light--matter interaction}
The standard expression for light matter interaction 
comes from minimal coupling ( $\mathbf{p} \rightarrow 
\mathbf{p} + e \mathbf{A} $ ), which gives the Hamiltonian
\begin{align}
\label{hamilton}
h =\frac{1}{2m} \mathbf{p}^2 + V + \frac{e}{m} \mathbf{p} \cdot \mathbf{A} + \frac{e^2}{2m} \cdot \mathbf{A}^2 
\end{align}
If the spatial dependence of $\mathbf{A}$ can be neglected the 
diamagnetic term, $\sim \mathbf{A}^2$,  can be removed through a unitary transformation 
of the wave function, $\Psi \rightarrow U \Psi$,
with 
\begin{align}
U=\exp\left[\frac{i}{\hbar}\left(e^2\int^tdt'A^2(t') \right)\right],
\nonumber
\end{align}
and the transformation of the time-dependent Schr{\"o}dinger equation, 
\begin{align}
\label{unitary}
U   h  U^{\dagger} U \Psi  = 
i \hbar U \frac{\partial}{\partial t} U^{\dagger} U \Psi.
\end{align} 
This gives one remaining interaction term, 
\begin{align}
\label{dipolv}
h^{\rm velocity}_I =\frac{e}{m} \mathbf{p} \cdot \mathbf{A},
\end{align}
which is usually referred to as the velocity gauge form.
A different   unitary transformation 
\begin{align}
U=\exp\left[\frac{i}{\hbar}\left( e \mathbf{r} \cdot \mathbf{A}(t) + e^2\int^tdt'A^2(t') \right)\right]
\nonumber
\end{align}
 can be employed to find  the alternative length gauge form
\begin{align}
\label{dipoll}
h^{\rm length}_I =e \mathbf{r} \cdot \mathbf{E},
\end{align} 
for details see e.g. Ref.~\cite{PhysRevA.76.023427}.
Here it is worth noting that in order to arrive at Eq.~(\ref{dipoll}) from 
Eq.~(\ref{unitary})
it is necessary to assume that the potential term in the Hamiltonian in Eq.~(\ref{hamilton}) commutes with $U$. This is obviously true for the Coulomb interaction with the nucleus, as well as between the 
electrons. 
 However, due to  the non-local nature of the Hartree-Fock exchange potential this is not the case within the Hartree-Fock approximation. Only by adding the RPAE class of many-body effects  can the invariance between the two forms be restored~\cite{lin:77}.
Close agreement between the two forms is  often considered a quality mark for more elaborate calculations. Since the agreement is trivial for any local potential it is considered a necessary, albeit not sufficient, property.

With linearly polarized light we may now write the transition matrix elements from Eq.~(\ref{onephoton2}) in length gauge as 
\begin{align}
\label{M1length}
M^{(1)} \left( q, \Omega, a \right) =  \langle q \mid  e z \mid a\rangle  E_{\Omega},
\end{align} 
or in velocity gauge as 
\begin{align}
\label{M1vel}
M^{(1)} \left( q, \Omega, a \right) =  \langle q \mid  \frac{e}{m} p_z \mid a\rangle  A_{\Omega}.
\end{align}
These non-correlated transition matrix elements can be chosen to be real in Eq.~(\ref{M1length}) and imaginary in  Eq.~(\ref{M1vel}) by use of real radial wave functions. 

Similarly the two-photon matrix element in Eq.~(\ref{twophoton}) can be written as 
\begin{align}
\label{M2}
M^{(2)}(q, \omega, \Omega, a ) =
\nonumber \\
\lim_{\xi \rightarrow 0^+}
\intsum{p} \frac{
\langle q \mid e z \mid p  \rangle \langle p  \mid    e z  \mid a \rangle}{\epsilon_a + \hbar \Omega -\epsilon_p + i \xi}  E_{\omega} E_{\Omega} .
\end{align}
in length gauge (and similarly with the $e p_z/m$ operator and vector potentials $A_\omega A_\Omega$ in velocity gauge).
An important difference compared to one-photon absorption is that the two-photon matrix element is intrinsically complex for above-threshold ionization, i.e. when the XUV photon energy exceeds the atomic binding energy, $\hbar\Omega>-\epsilon_a>0$.

The atomic contribution to the quantum beating of the side band at energy, $\epsilon_{q} = 2n \hbar \omega + \epsilon_a$,  in a RABBIT experiment is the phase difference between  the  quantum path where the XUV harmonic $\hbar \Omega_> = \left(2n+1\right)\hbar \omega $ is absorbed and an  IR-photon is emitted and that where both an XUV harmonic, now of energy $\hbar \Omega_< = \left(2n-1\right)\hbar \omega$, and an IR-photon is absorbed. Eq.~(\ref{M2}) shows the most important path,   but  contributions will also come from the reversed time-order where the IR photons are exchanged before absorption of any XUV photon.  For this latter path there is in the general case no on-shell intermediate state that can contribute. It is thus assumed to be of less importance  and is consequently often neglected. While this is a justified approximation for calculations in length gauge the situation is very different in velocity gauge as we will see below.

\subsection{The time delay}
\label{timedelay}
Following the usual RABBIT formalism \cite{DahlstromJPB2014}, we construct the phase shift of photoelectrons that take two different quantum paths leading to the same final state with momentum along the common polarization axis of the fields, $\mathbf{\hat z}$, as 
\begin{align}
\label{phi2}
\phi_\mathrm{a}
&=\arg \left(\sum_{\ell}M_\mathrm{a}(\ell)
e^{i\left( - \ell \frac{\pi}{2} + \eta_{Z,k,\ell}\right) } 
Y_{\ell,0}(\mathbf{\hat z}) 
\right) 
\nonumber \\
\phi_\mathrm{e}
&= \arg \left(\sum_{\ell} M_\mathrm{e}(\ell)
e^{i \left( - \ell \frac{\pi}{2} + \eta_{Z,k,\ell} \right) }
Y_{\ell,0}(\mathbf{\hat z}) 
\right).
\end{align}
At this emission angle only the zero magnetic quantum number contributes to the ionization process, $m_a=0$. We use the following short-hand notation,
\begin{align}
M_\mathrm{a}(\ell) &=  M^{(2)}(q, \omega, \Omega_<, a ), \nonumber \\
M_\mathrm{e}(\ell) &=  M^{(2)}(q, -\omega, \Omega_>, a ) \nonumber \\
\eta_{Z,k,\ell} &= \sigma_{Z,k,\ell} + \delta_{k,\ell}, \nonumber
\end{align}
where subscripts $\mathrm{a}$ and $\mathrm{e}$ stand for IR absorption and emission, respectively 
(do not confuse the subscript $\mathrm{a}$ with the quantum number label $a$ for the initial atomic state),  
where $M_\mathrm{a/e}(\ell)$ depend on angular momentum $\ell$ of the final $q$--state.
The atomic delay for emission along $\mathbf{\hat z}$ can be calculated for sideband $2n$ as 
\begin{align}
\label{atomicdelay}
\tau_{A}=\frac{\phi_\mathrm{e}-\phi_\mathrm{a}}{2 \omega}.
\end{align}
Similarly the one-photon phase shifts of the photoelectron in the $\mathbf{\hat z}$ direction are
\begin{align}
\label{deltaphiWigner}
\phi_>
&=
 \arg 
\left( \sum_{\ell} 
M_>
(\ell)
e^{i 
\left( - \ell \frac{\pi}{2} + \eta_{Z,k_>,\ell} 
\right) }
 Y_{\ell,0}(\mathbf{\hat z}) \right) 
\nonumber \\
\phi_<
&=  
\arg   
\left( \sum_{\ell}
M_<
(\ell)
e^{i 
\left( - \ell \frac{\pi}{2} + \eta_{Z,k_<,\ell}
\right) } Y_{\ell,0}(\mathbf{\hat z})
\right),
\end{align}
where we use short-hand notation for the one-photon matrix elements, 
$M_{>/<}(\ell)\equiv M^{(1)}(q_{>/<},\Omega_{>/<},a)$, 
with final  photoelectron wave number $k_{>/<}$ and angular momentum $\ell$, 
after absorption of a photon with angular frequency $\Omega_{>/<}$. 
Eq.~(\ref{deltaphiWigner}) can be used to compute the Wigner-like delay at sideband $2n$ along $\mathbf{\hat z}$ as
\begin{align}
\label{wignerdelay}
\tau_W= \frac{
\phi_> 
-
\phi_< 
}{2 \omega}. 
\end{align}
We point out that the definition of Wigner delay using Eq.~(\ref{wignerdelay}) breaks down at resonances that typically have large phase variations over the photon energy of the probe field \cite{DahlstromJPB2012}. 
In the following we refer to the quantity $\tau_A-\tau_W$ as the delay {\it difference} induced by the laser field in RABBIT. We make a distinction between this ``exact'' delay difference and the approximate continuum-continuum delay that can be derived using asymptotic continuum functions, $\tau_{CC}\approx\tau_A-\tau_W$ \cite{DahlstromCP2013}. 

\section{Method} 
\label{method}
While the RPAE-approximation has been used to include electron correlation effects for the interaction with the ionizing XUV photon, the subsequent above-threshold interaction with the IR field has been limited to a static atomic interaction in our earlier studies~\cite{dahlstrom:12,DahlstromJPB2014,negativeiondelay:2017}. 
Here we discuss, in some detail,  how this approximation can be lifted. 
The calculations are performed with a 
basis set obtained through diagonalization of  effective one-particle Hamiltonians in a radial primitive basis of B-splines~\cite{deboor}, in a spherical box.  For each angular momentum $\ell$ this 
 one-particle Hamiltonian reads:
\begin{align}
\label{oneparticle}
h_{0}^{\ell}\left(r\right) = -\frac{\hbar^2}{2m}\frac{\partial^2}{\partial r^2} +
\frac{\hbar^2}{2m}\frac{\ell \left(\ell+1\right)}{r^2} 
\nonumber \\ -\frac{e^2}{4\pi \epsilon_0}\frac{Z}{r} + u_{{\rm HF}} + u_{{\rm proj}}.
\end{align}
It includes the (non-local) Hartree-Fock potential (HF), $u_{{\rm HF}}$,  for the closed shell with $N$ electrons and a correction, $u_{{\rm proj}}$ (also non-local). The latter is called a projected potential (for the  explicit form see Sec.~\ref{methodA} below) and it  ensures that any excited electron feels an approximate long-range potential with $N-1$ electrons remaining on the target. Since it is projected on virtual states  it does not affect the occupied HF orbitals. The projected potential allows us to include some effects already in the basis set, that would otherwise be treated perturbatively through the RPAE-iterations. The eigenstates to $h_{0}^\ell$ form an orthonormal basis with eigenenergies  $\epsilon_i$ that is used for the description of the occupied orbitals, but it is also used to span the virtual space of the photoelectron.

\begin{figure*}
\includegraphics[width=0.98\textwidth]{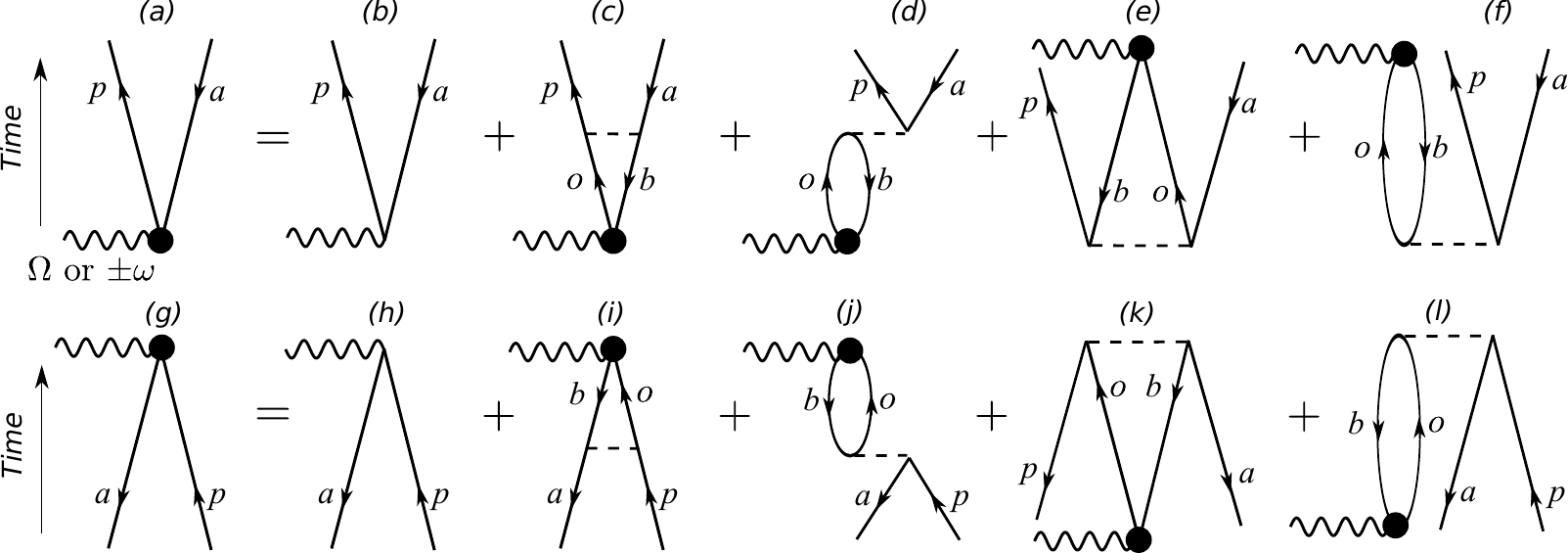}
\caption{RPAE for the many-body screening  of the photon interaction. (a) and (g) are forward and backward propagation, respectively, where the sphere indicates the correlated interaction to infinite order.
\label{fig:rpae}}
\end{figure*}

We start with writing the dipole interaction between one electron and the electromagnetic fields as 
\begin{align}
 h_I=& 
 \lim_{\xi\rightarrow 0^+} 2\sum_j d_{\Omega_j} \cos(\Omega_j t)e^{\xi t}  \nonumber  \\
=& \lim_{\xi\rightarrow 0^+} \sum_j d_{\Omega_j} \sum_{\sigma=\pm1}
  \exp[(-i\sigma\Omega_j+\xi)t], 
\end{align}
where $d_{\Omega_j}$ is a time-independent operator that describe the coupling of the atom to the field with angular frequency $\Omega_j$ and $\xi$ is used to set the outgoing boundary condition for the interaction.   With linearly polarized light the length gauge expression is 
 $d_{\Omega_j} =e z E_{\Omega_j}$, where $ez$ is the dipole operator component along the field polarization.  Consider now an electron in occupied orbital $\mid a\rangle$, i.e. in an eigenstate to the one-particle Hamiltonian in Eq.~(\ref{oneparticle}). When it absorbs ($\sigma=1$), or  emits ($\sigma=-1$) photons it will acquire correction terms to its wave function of the type:
\begin{align}
\label{aplusdelta}
\mid\psi_a \left(t \right)\rangle =
\mid a \rangle e^{-i \epsilon_a t/\hbar} + 
\sum_{j}\sum_{\sigma=\pm1}
\mid \rho^{(1:\sigma)}_{\Omega_j,a} \rangle 
e^{-i \left(\sigma \Omega_j+\epsilon_a/\hbar\right)t}   \nonumber \\
+  \sum_{j,j'} \sum_{\sigma=\pm1} \sum_{\sigma'=\pm1}\mid \rho^{(2:\sigma',\sigma)}_{\Omega_{j'},\Omega_{j},a} \rangle  
e^{-i \left( \sigma' \Omega_{j'}+\sigma  \Omega_j+\epsilon_a/\hbar \right)t}   + \ldots,
\end{align}
where the superscripts and subscripts label sequences of interactions with photons (signs and angular frequencies) by joint increasing primes. Expressions for the correction terms can be found through the time-dependent Schr{\"o}dinger equation
\begin{align}
\label{TDSE}
\left(i\hbar \frac{\partial}{\partial t} - h_{0}^{\ell}\right)
\mid\psi_a \left(t \right)\rangle
= h_I \mid\psi_a \left(t \right)\rangle,
\end{align}
by collecting the contributions that scale linearly with the field $E_{\Omega_j}$ and oscillate with
$\exp[-i(\sigma\Omega_j+\epsilon_a/\hbar)t]$ as
\begin{align}
\label{firstorder_a}
\left(\sigma \hbar \Omega_j + \epsilon_a - h_{0}^{\ell} \right) 
 \mid \rho^{(1:\sigma)}_{\Omega_j,a}\rangle = 
d_{\Omega_j}\mid a \rangle.
\end{align}
For a single electron case the desired one-photon correction to the wave function is simply obtained from Eq.~(\ref{firstorder_a}), which we call the one-electron first-order perturbed wave function, 
\begin{align}
\label{rho0}
\mid \rho^{(1:\sigma)}_{0, \Omega_j,a} \rangle  
=
\sum_{p}\frac{\mid p \rangle \langle  p \mid d_{\Omega_j} \mid a \rangle}{
\sigma  \hbar \Omega_j+\epsilon_a - \epsilon_p }
,
\end{align}
where the sum over $p$ runs over all states (including also the continuum). For a many-electron system, however, there are more effects to consider.  The starting point is then a Slater determinant $\mid\{ab \ldots n\}\rangle$ (where curly brackets denote anti-symmetrization, $\mid\{ab \ldots n\}\rangle\equiv(\mid ab  \ldots n \rangle-\mid ba\ldots n\rangle \ldots )/\sqrt{n!}$   ). The field-corrected  wave function will also be a Slater determinant, but now  the orbitals are  as given by  Eq.~\ref{aplusdelta}, i.e
\begin{align}
\label{aplusdeltaMB}
\mid\Psi \left(t \right)\rangle = \mid \left\{\psi_a\left(t \right), \psi_b\left(t \right),\psi_c\left(t \right),\psi_d\left(t \right),\ldots \right\} \rangle
\end{align}
Since the interaction with the other electrons is accounted for by the Hartree-Fock potential the possible changes in it due to the interaction with the electromagnetic field have to be considered. We will return to this question below, but here we mark that the sum on $p$ in Eq.~(\ref{rho0}) is restricted to {\it unoccupied} states for the many-electron case (below the sums will be marked $exc$ to include only these ``excited'' states). This is simply what is expected from the Pauli exclusion principle. Alternatively, the restriction of $p$ to excited states can be understood from Eq.~(\ref{aplusdeltaMB}) where a $\sigma=+1$ excitation of orbital $a$ into $b$, as given  in Eq.~(\ref{rho0}), will cancel the $\sigma=-1$ excitation of orbital $b$ into $a$ and vice verse.

We are interested in photoionization processes that happen when $\sigma \hbar \Omega + \epsilon_a >0$. This implies that there is a pole in the denominator of Eq.~(\ref{rho0}) that must be treated with the proper boundary condition and continuum integration. An efficient way to do this is to use exterior complex scaling (ECS) of the radial coordinate,   
\begin{align}
\label{complexrotation}
r \rightarrow
\Bigg \{
\begin{array}{lr}
r, & 0 < r  < R_C\\
R_C + \left(r-R_C\right) e^{i \varphi}, & r > R_C, 
\end{array}
\end{align}
which enforces the outgoing boundary condition for the unbound states.
The eigenenergies of the orbitals are complex in general using ECS, and it has the advantage that the integration over the continuum is effectively performed 
by a sum over a discretized representation of all excited states, $p$, as written in Eq.~(\ref{rho0}). 

The terms in Eq.~(\ref{TDSE}) proportional to the product $E_{\Omega_j'} E_{\Omega_{j}}$ that oscillates with 
$\exp[-i(\sigma'\Omega_{j'}+\sigma\Omega_j+\epsilon_a/\hbar)t]$ are
\begin{align}
\label{secondorder_a}
\left(\sigma' \hbar \Omega_{j'}+\sigma \hbar \Omega_j+\epsilon_a - h_{0}^{\ell} \right) 
 \mid\rho^{(2:\sigma',\sigma)}_{\Omega_{j'},\Omega_j,a}\rangle = 
d_{\Omega_{j'}} \mid \rho^{(1:\sigma)}_{\Omega_j,a} \rangle,
\end{align}
for the case where the $\sigma'$ interaction with frequency $\Omega_{j'}$ happens after the $\sigma$ interaction with frequency $\Omega_{j}$. 
The second-order correction for a single electron are 
\begin{align}
\label{rho00}
\mid \rho^{(2:\sigma',\sigma)}_{0,\Omega_{j'},\Omega_j,a} \rangle  =
\sum_{p}\frac{\mid p \rangle \langle  p \mid d_{\Omega_{j'}} \mid \rho^{(1:\sigma)}_{0, \Omega_j,a} \rangle}{
\sigma'\hbar \Omega_{j'}+\sigma \hbar \Omega_{j}+\epsilon_a - \epsilon_p },
\end{align}
which simply builds on the first-order correction. 
Next, we need to define a notation for the corrections that includes all possible time orders by writing a square bracket around the signs, $[\sigma',\sigma]$, and frequencies, $[\Omega_{j'},\Omega_j]$, to include all {\it joint} permutations of primes on signs and frequencies. The  second-order correction for a single electron with summed time orders is simply 
\begin{align}
\label{rho00TO}
\mid \rho^{(2:[\sigma',\sigma])}_{[\Omega_{j'},\Omega_j],a} \rangle  =
\sum_{p}\frac{\mid p \rangle \langle  p \mid }{
\sigma'\hbar \Omega_{j'}+\sigma \hbar \Omega_{j}+\epsilon_a - \epsilon_p }
\nonumber \\
\times\left(
d_{\Omega_{j'}} \mid \rho^{(1:\sigma)}_{0, \Omega_j,a} \rangle
+
d_{\Omega_{j}} \mid \rho^{(1:\sigma')}_{0, \Omega_{j'},a} \rangle\right).
\end{align}

\subsection{One-photon RPAE}
\label{methodA}
The many-body response to the interaction with the photon is neglected in Eq.~(\ref{rho0}), but the bulk of these effects can be added through the RPAE method~\cite{Amusia1990}, where certain sub-classes of many-body effects are included  through the iterative solution of the equations for the coupled channels.
Another name for  RPAE is time-dependent Hartree-Fock~\cite{jamieson:70}, and we will here use that point of view to derive the expressions we need.  With the HF-approximation  each  orbital is   described as moving in an average potential from the other orbitals, and its  matrix element between any orbitals $m, n$ (occupied or unoccupied),  is:  
\begin{align}
\langle  m \mid  u_{HF} \mid n \rangle = \sum_b^{core} \langle \{ m b \} \mid  V_{12} \mid \{n b\}\rangle, 
\label{HF}
\end{align}  
where the Coulomb interaction is given by 
\begin{align}
V_{12} = \frac{e^2}{4\pi \epsilon_0} \frac{1}{r_{12}} = \frac{e^2}{4\pi \epsilon_0}
\sum_{K=0}^{\infty} \frac{r_<^K}{r_>^{K+1}}  \mathbf{C}^K \left( 1 \right) \cdot \mathbf{C}^K\left( 2 \right).
\end{align}

Our starting point is a Slater determinant constructed from orbitals that are  solutions to Eq.~(\ref{oneparticle}), with the 
Hartree-Fock potential defined as in Eq.~(\ref{HF}).
When the electrons interact with the field and acquire  perturbations according to Eq.~(\ref{aplusdelta}) the potential itself will  change, $u_{HF} \rightarrow u_{HF} + \delta u$. This  gives  rise to  additional paths for orbital $a$ to absorb or emit one photon with phase factor $\exp[-i\sigma\Omega_jt]$. In Eq.~(\ref{HF}) we replace $b \rightarrow b + \rho_{\Omega_j,b}^{(1:\sigma)}$, let the potential work on orbital $a$, and identify new terms to the excited states, $p$, that are linear in the electric field and oscillate with
$\exp\left[
-i(\sigma\Omega_j+\epsilon_a/\hbar)t
\right]$
as 
\begin{align}
\langle  p \mid  \delta u_{\Omega_j}^{(1:\sigma)} \mid a \rangle =&
\sum_b^{core} \left[ 
\langle \{p b \} \mid  V_{12} \mid  \{ a \rho_{\Omega_j,b}^{(1:\sigma)}\} \rangle \right.
\nonumber \\
+& 
\left.
\langle \{ p \rho_{\Omega_j, b}^{(1:-\sigma)} \}\mid  V_{12} \mid \left\{ a b \right\} \rangle
\right].
\label{diagrams}
\end{align}
In the case of absorption of a photon, $\sigma=1$, this implies that the second term in Eq.~(\ref{diagrams}) is generated using a perturbed wave function that describes virtual emission of a photon, $\sigma=-1$. Adding Eq.~(\ref{diagrams}) as an additional source term to the right-hand side of Eq.~(\ref{firstorder_a}) leads to coupled equations for the correlated perturbed wave functions for absorption and emission of a photon,    
\begin{align}
\label{eqrpae1}
\left(\sigma \hbar \Omega_j +\epsilon_a - h \right) \mid  \rho^{(1:\sigma)}_{\Omega_j, a}\rangle  
=\sum_{p}^{exc} 
\mid p\rangle \langle  p \mid 
\left( 
d_{\Omega_j} + \delta u_{\Omega_j}^{(1:\sigma)} 
\right)
\mid a \rangle
.
\end{align}
Use of Eqs.~(\ref{rho0}) and (\ref{diagrams}) leads to the final expression
\begin{align}
\label{RPA}
\mid \rho^{(1:\sigma)}_{\Omega_j,a} \rangle
= \mid \rho^{(1:\sigma)}_{0,\Omega_j,a} \rangle 
-\sum_p^{exc} \frac{\mid p \rangle}{\sigma \hbar \Omega_j + \epsilon_a -\epsilon_p} \nonumber \\
\times \Bigg(  \sum_b^{core}
\Bigg[
\langle b p \mid V_{12} \mid a \,  \rho^{(1:\sigma)}_{\Omega_j,b} \rangle
- \langle b \,  p \mid V_{12} \mid   \rho^{(1:\sigma)}_{\Omega_j,b} \,  a \rangle \nonumber \\
+ \langle \rho^{(1:-\sigma)}_{\Omega_j,b} \,  p \mid V_{12} \mid    a b \rangle
- \langle p  \, \rho^{(1:-\sigma)}_{\Omega_j,b}  \mid V_{12} \mid    a b \rangle \Bigg] \nonumber \\
-  \langle p  \mid u_{{\rm proj}} \mid \rho^{ (1:\sigma)}_{\Omega_j,a}\rangle
\Bigg),
\end{align}
where the exchange interactions are written out explicitly. 
The upper part of Fig.~\ref{fig:rpae} shows the  Goldstone diagrams for 
$\mid \rho^{(1:\sigma)}_{\Omega_j,a}\rangle$, 
where Fig.~\ref{fig:rpae}~(b) is the uncorrelated absorption of a photon $\Omega_j$, corresponding to the first term on the right-hand side of Eq.~(\ref{RPA}). 
Figs.~\ref{fig:rpae}~(c) and (d) account for the electron--hole interaction in forward propagation, corresponding to the second and third terms,
while Figs.~\ref{fig:rpae}~(e) and (f) account for ground-state correlation effects, corresponding to the forth and fifth terms on the right-hand side of Eq.~(\ref{RPA}).  
The last term in Eq.~(\ref{RPA}) removes the projected potential, introduced in Eq.~(\ref{oneparticle}), 
which we take to be the monopole interaction with a given hole $c$,  
\begin{align}
\label{proj}
 u_{{\rm proj}}  =  -\frac{e^2}{4\pi\epsilon_0} \sum_{r,s}^{exc} \mid r \rangle \langle r c \mid \frac{1}{r_>} \mid s c \rangle \langle s  \mid. 
\end{align} 
It will cancel the corresponding part of $\delta u$ of Fig.~\ref{fig:rpae}~(c) and (i) with $a=b=c$ and $K=0$. When converged, the iterative procedure gives the same results  if the projected potential is used or not, but the convergence is often much improved in the latter case, especially close to  ionization thresholds.

\subsection{Two-photon RPAE}
We now derive the interaction with two photons for the multi-electron case. 
The second interaction with the field can stimulate either the excited electron or the remaining hole from the first interaction. The latter effect
arise when the staring point is a Slater determinant and  the corrected wave function is of the form given in Eq.~(\ref{aplusdeltaMB}). The net result is a   coupling of the wave functions associated different holes in Eq.~(\ref{aplusdelta}), by the hole-hole dipole interaction in the source term of Eq.~(\ref{TDSE}). Collecting the terms proportional to $E_{\Omega_{j'}} E_{\Omega_j}$ 
from Eq.~(\ref{TDSE}) that oscillate with $\exp[-i(\sigma'\Omega_{j'}+\sigma\Omega_{j}+\epsilon_a/\hbar)t]$, we write
\begin{align}
\label{rho02}
\left( \sigma'\hbar \Omega_{j'} + \sigma\hbar \Omega_j + \epsilon_a - h_{0}^{\ell} \right)
\mid \rho^{(2:[\sigma',\sigma])}_{0,[\Omega_{j'},\Omega_j],a}\rangle =
\nonumber \\ 
\sum_p^{exc}
\mid p \rangle \langle p \mid 
\left( 
d_{\Omega_{j'}}\mid \rho^{ (1:\sigma)}_{\Omega_j,a}\rangle 
+
d_{\Omega_{j}}\mid \rho^{ (1:\sigma')}_{\Omega_{j'},a}\rangle 
\right)
\nonumber \\ 
-\sum_c^{core}
\left(
\langle c \mid d_{\Omega_{j'}}\mid a \rangle
\mid \rho^{(1:\sigma)}_{\Omega_j,c}\rangle 
 +
\langle c \mid d_{\Omega_{j}}\mid a \rangle
\mid \rho^{(1:\sigma')}_{\Omega_{j'},c}\rangle 
\right),
\end{align}
where the source terms on the right-hand side contain both time orders. 
In Eq.~(\ref{rho02}) the second line accounts for the interaction with the excited electron, 
Fig.~\ref{fig:M2full}~(a) and (b), 
while the third line accounts for hole transfer from another orbital, Fig.~\ref{fig:M2full}~(c) and (d). The minus on the third line comes from Wick's theorem, which is evaluated using the Goldstone rules associated with the diagrams in Fig.~\ref{fig:M2full} \cite{mbpt}.    

The next step is to consider the  many-body response. Second-order corrections to the Hartree-Fock potential can generate terms proportional to $E_{\Omega_{j'}} E_{\Omega_{j}}$. By letting 
$b \rightarrow b + \rho_{\Omega_{j}, b}^{(1:\sigma)} + \rho^{(2:[\sigma',\sigma])}_{[\Omega_{j'},\Omega_j],b}...$ in Eq.(\ref{HF}), and collecting the terms that oscillate with $\exp[-i(\sigma'\Omega_{j'}+\sigma\Omega_{j}+\epsilon_a/\hbar)t]$ we arrive at:
\begin{align}
\langle  p \mid  \delta u_{[\Omega_{j'},\Omega_j]}^{(2:[\sigma',\sigma])} \mid a \rangle =
\nonumber \\
\sum_b^{core} \Bigg(
\langle 
\{ p b \} \mid V_{12} \mid\{a\rho_{[\Omega_{j'},\Omega_j],b}^{(2:[\sigma',\sigma])} \}\rangle 
\nonumber \\  +
\langle \{ p \rho_{[\Omega_{j'},\Omega_j],b}^{(2:[-\sigma',-\sigma])} \} \mid V_{12} \mid \{ a b  \} \rangle
\nonumber \\  +  
\langle \{p \rho_{\Omega_{j'},b}^{(1:-\sigma')} \}\mid V_{12} \mid \{ a \rho_{\Omega_{j}, b}^{(1:\sigma)} \}\rangle 
\nonumber  \\
\langle \{p \rho_{\Omega_{j},b}^{(1:-\sigma)} \}\mid V_{12} \mid \{ a \rho_{\Omega_{j'}, b}^{(1:\sigma')} \}\rangle 
\Bigg)
\label{VHF2}
\end{align}
The forward propagating ($\sigma=1$), direct contributions from lines two and three are depicted in Fig.~\ref{fig:M2full}~(k-l), and those from lines four and five in Fig.~\ref{fig:M2full}~(i) (only one of the two time-orders is shown).
Another set of contributions, that will have the right oscillations, are
the first order corrections from the Hartree-Fock potential when they,  just as the dipole operator in Eq.~(\ref{rho02}), work on   the corrected wave functions. This gives corrections
\begin{align}
\label{deludelrho}
\langle  p \mid  \delta u_{\Omega_{j'}}^{(1:\sigma')} \mid  \rho_{\Omega_{j},a}^{(1:\sigma)} \rangle =&
\sum_b^{core} \left[ 
\langle \{p b \} \mid  V_{12} \mid  \{ \rho_{\Omega_{j},a}^{(1:\sigma)} \rho_{\Omega_{j'},b}^{(1:\sigma')}\} \rangle \right.
\nonumber \\
+& 
\left.
\langle \{ p \rho_{\Omega_{j'}, b}^{(1:-\sigma')} \}\mid  V_{12} \mid \left\{ \rho_{\Omega_{j},a}^{(1:\sigma)} b \right\} \rangle
\right],
\end{align}
for which  the direct contributions are depicted in Fig~\ref{fig:M2full}~(e) and (g), and also
\begin{align}
\label{deludelrhoc}
\langle  c \mid  \delta u_{\Omega_{j'}}^{(1:\sigma')} \mid  a\rangle \,  \rho_{\Omega_{j},c}^{(1:\sigma)}  =&
\sum_b^{core} \left[ 
\langle \{c b \} \mid  V_{12} \mid  \{  a \rho_{\Omega_{j'},b}^{(1:\sigma')}\} \rangle 
\rho_{\Omega_{j},c}^{(1:\sigma)}
\right.
\nonumber \\
+& 
\left.
\langle \{ c \rho_{\Omega_{j'}, b}^{(1:-\sigma')} \}\mid  V_{12} \mid \left\{a b \right\} \rangle
 \rho_{\Omega_{j},c}^{(1:\sigma)}
\right],
\end{align}
where the direct contributions are  depicted in Fig~\ref{fig:M2full}~(f) and (h).
Note though that in both Eq.~(\ref{deludelrho}) and Eq.~(\ref{deludelrhoc}) 
the case when $j'$ and $j$ are interchanged is to be added. 
Finally, there are second-order corrections that stem from  the fact that the expression in Eq.~(\ref{aplusdelta}) uses 
{\em intermediate normalization}, which means that the occupied orbitals, 
$\mid a \rangle$, 
are normalized and orthogonal to the corrections, 
$\mid \rho^{(1:\sigma)}_{\Omega_j,a}\rangle, \,...$, 
while $\mid\psi_a\rangle$ is neither normalized nor orthogonal to $\mid\psi_b\rangle$. These corrections for the second-order interaction depend on the inner-product of the first-order corrections to the wave functions,  
\begin{align}
\langle p \mid N^{(2:[\sigma',\sigma])}_{[\Omega_{j'},\Omega_j]} \mid a\rangle = 
-\sum_{b,c}^{core}
 \langle \left\{p b \right\}\mid V_{12} \mid \left\{ a c \right\}\rangle 
\nonumber \\
\times 
\left(
 \langle \rho_{\Omega_{j'},c}^{(1:-\sigma')}\mid \rho_{\Omega_{j},b}^{(1:\sigma)} \rangle +
 \langle \rho_{\Omega_{j},c}^{(1:-\sigma)}\mid \rho_{\Omega_{j'},b}^{(1:\sigma')} \rangle 
\right).
\label{N2}
\end{align}
Again the direct contribution  is depicted in Fig.~\ref{fig:M2full}~(j) for one of the time-orders. The contributions from Eqs.~(\ref{VHF2} - \ref{N2}) should now be added as source terms 
to Eq.~(\ref{rho02}) and we can write down the equation for the second order correction including the many-body response:
\begin{align}
\left( \sigma'\hbar \Omega_{j'} + \sigma\hbar \Omega_j + \epsilon_a - h_{0}^{\ell} \right)
\mid \rho^{(2:[\sigma',\sigma])}_{[\Omega_{j'},\Omega_j],a}\rangle =
\nonumber \\ 
=
\sum_p^{exc}
\mid p \rangle \langle p \mid 
\left[
%
\left(
 \delta u_{[\Omega_{j'},\Omega_j]}^{(2:[\sigma',\sigma])}
+  N^{(2:[\sigma',\sigma])}_{[\Omega_{j'},\Omega_j]}
\right) \mid a \rangle  
 \right. \nonumber \\ \left.
+ 
\left( \delta u_{\Omega_{j'}}^{(1:\sigma')} +
d_{\Omega_{j'}} \right) \mid \rho^{ (1:\sigma)}_{\Omega_j,a}\rangle +  
  \left( \delta u_{\Omega_{j}}^{(1:\sigma)} + d_{\Omega_{j}} \right) \mid \rho^{ (1:\sigma')}_{\Omega_{j'},a}\rangle 
\right. \nonumber \\ \left.
- u_{{\rm proj}} \mid  \rho^{(2:[\sigma',\sigma])}_{[\Omega_{j'},\Omega_j],a}\rangle 
\right]
\nonumber \\ 
- \sum_c^{core}
 \langle c \mid
\left[
\left(  d_{\Omega_{j'}} +  \delta u_{\Omega_{j'}}^{(1:\sigma')} \right)\mid a \rangle
\mid \rho^{(1:\sigma)}_{\Omega_j,c}\rangle 
\right.  \nonumber \\ \left. 
 + \left( d_{\Omega_{j}}  + \delta u_{\Omega_{j}}^{(1:\sigma)} \right)\mid a \rangle
\mid \rho^{(1:\sigma')}_{\Omega_{j'},c}\rangle 
\right].
\label{M2Eq}
\end{align}

The term with $-u_\mathrm{proj}$ compensates for the projected potential, which, as mentioned above, is important only for numerical convergence.

\subsection{Calculation of two-photon matrix elements}
\label{methodC}
\begin{figure*}
\includegraphics[width=0.90\textwidth]{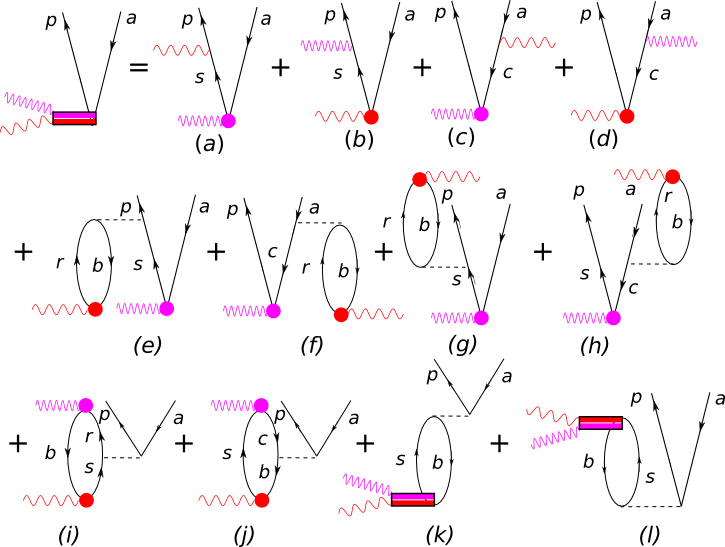}
\caption{Goldstone diagrams illustrating the contributions to the forward propagating two-photon RPAE perturbed wave function, $ \mid \rho_\mathrm{a/e}\rangle$ in~Eqs.~(\ref{M2Eq}) - (\ref{rho2plusminus}). Only direct diagrams are shown. For diagrams  (g-j) there are also contributions with the order of the two photons interchanged. There is a similar equation for the backward propagating diagrams needed to evaluate diagram (l). The calculations include the full set of diagrams including the exchange versions and the time-orders omitted from the illustration.
\label{fig:M2full}}
\end{figure*}

For a RABBIT calculation with photoelectron energy $2n\hbar\omega+\epsilon_a$,   
we need two specific second-order correlated perturbed wavefunctions for orbital $a$ from Eq.~(\ref{M2Eq}),  
\begin{align}
\label{rho2plusminus}
\mid\rho^{(2:[\pm,+])}_{[\omega,(2n\mp1)\omega],a}\rangle \equiv \, \mid \rho_\mathrm{a/e}\rangle, 
\end{align}
that include absorption  of a smaller XUV photon and absorption,  ($\mathrm{a}$), of a laser photon,
as well as absorption of a larger XUV photon with emission,  ($\mathrm{e}$), of a laser photon, denoted $\mid\rho_\mathrm{a/e}\rangle$ for brevity.  Given $\mid \rho_\mathrm{a/e}\rangle$ we may directly extract the two-photon matrix elements needed for the 
calculation of the atomic delay, c.f.~Sec.\ref{timedelay}. However, due to the on-shell above threshold contributions to the diagram in  Fig.~\ref{fig:M2full}~(a), the construction of  $\mid\rho_\mathrm{a/e}\rangle$ for the time-order where the XUV pulse is absorbed first involves an integration over a double pole and  is not trivial. To circumvent this problem we first calculate
the two-photon matrix element for the diagrams in Fig.~\ref{fig:M2full}~(a-d) and treat the additional corrections to $\mid\rho_\mathrm{a/e}\rangle$ separately. The different steps are detailed below.

The contributions from  Fig.~\ref{fig:M2full}~(a-d) 
are  calculated directly from the first order corrections  $|\rho^{(1:+)}_{(2n\mp1)\omega,a}\rangle$. 
In length gauge, the diagrams in Fig.~\ref{fig:M2full}~(a) and (c) amount to:
\begin{align}
\label{TO1}
M^{\mathrm{TO:1}}_\textrm{a/e} &= \Big( \langle q \mid ez \mid  \rho^{(1:+)}_{(2n\mp1)\omega,a} 
\rangle  \\ \nonumber & -
 \sum_c^{core} \langle q \mid   \rho^{(1:+)}_{(2n\mp1)\omega,c} 
\rangle \langle c  \mid  ez \mid a \rangle
\Big)  / E_{(2n\mp1)\omega},
\end{align}
where TO:1 stands for first time order, while the diagrams in Fig.~\ref{fig:M2full}~(b) and (d) amount to
\begin{align}
\label{TO2}
M^\mathrm{TO:2}_\textrm{a/e} &= \Big(  \langle q \mid ez \mid  \rho^{(1:\pm)}_{\omega,a} \rangle 
\nonumber \\
&- \sum_c^{core} \langle q \mid  \rho^{(1:\pm)}_{\omega,c} \rangle \langle c  \mid  ez \mid a \rangle
 \Big)   /E_{\omega},
\end{align}
where TO:2 stands for the second time order. The  final state $q$ is here an eigenstate to the effective one-particle Hamiltonian at the sideband kinetic energy $\epsilon = \epsilon_a + 2n\hbar\omega$. As described in Ref.~\cite{DahlstromJPB2014,negativeiondelay:2017} the numerical representation of the radial part of $\mid q \rangle$,   denoted $P_{q}(r)$,  is a solution of  
\begin{align}
h_{\ell} P_{q}(r) = \epsilon P_{q}(r),
\end{align}
which can be reformulated as a system of linear equations  
for the unknown coefficients $c_i$ when expanded in B-splines
\begin{align}
 P_{q}(r) = \sum_i c_i B_i(r). 
\end{align}
For the case in Fig.~\ref{fig:M2full}~(a), where the first photon is of an XUV-wavelength causing  ionization, and the second integral is between two continuum states, the integral in  Eq.~(\ref{TO1}) will not converge for any finite interval on the real axis. The integration  is instead  performed numerically out to a distance far outside the atomic core, but within the unscaled region ($a_0\ll r<R_C$), while the final part of the integral is carried out using analytical Coulomb waves along the imaginary $r$-axis as described in Ref.~\cite{DahlstromJPB2014}. The numerical stability is monitored by comparison of  different ``break points'' between the numerical and analytical descriptions. The integrals in Eq.~(\ref{TO2}), on the other hand, converge inside the numerical box since the IR-field can only induce a localized correction to the wave function. 

The diagrams in Fig.~\ref{fig:M2full}~(e--j), and their exchange/switched time-order counterparts,  can all be calculated by connecting converged first-order corrections with a single Coulomb interaction. Finally the diagrams in Fig.~\ref{fig:M2full}~(k--l) are found in an iterative procedure following that for the first order correction, Eq.~(\ref{RPA}). With the use of the projected potential, Eq.~(\ref{proj}), all monopole terms are removed from the iterative procedure and the integral over the remaining Coulomb interaction does indeed converge on a finite interval. Therefore, it can be treated numerically inside the computational box. Separating the two-photon perturbed wave function in the lowest order contributions [Fig.~\ref{fig:M2full}~(a--d)] and the rest,
\begin{align}
|\rho_\mathrm{a/e}\rangle = 
|\rho_\mathrm{a/e}^{(0)}\rangle +
|\delta \rho_\mathrm{a/e}\rangle 
\nonumber
\end{align}
and the the remaining contribution to the two-photon matrix element, $\delta M_\mathrm{a/e}$, can be deduced directly from $|\delta \rho_\mathrm{a/e}\rangle$,  giving the final result:
\begin{align}
M_\mathrm{a/e} = M_\mathrm{a/e}^\mathrm{TO:1}+
M_\mathrm{a/e}^\mathrm{TO:2} + \delta M_\mathrm{a/e}.
\end{align}

Of the four contributions in Eq.~(\ref{TO1} - \ref{TO2}) it is natural to assume that the first term in Eq.~(\ref{TO1}), Fig.~\ref{fig:M2full}~(a), is by far the dominating because it suffers from a zero in the denominator of the perturbed wavefunction. In contrast,  Fig.~\ref{fig:M2full}~(b--d) are all connected with rather large denominators and should be small in general.  
The concept of cc-delays~\cite{DahlstromJPB2012,DahlstromCP2013},  where a photoelectron interact with a laser field after photoionization, derives from the assumption that the total two-photon process is well described by Fig.~\ref{fig:M2full}~(a) with use of a suitable long range potential, such as the projected potential in Eq.~(\ref{proj}) \cite{dahlstrom:12,DahlstromJPB2014}. 
Here we will show that this assumptions is close to the truth for calculations in length gauge, but wrong in velocity gauge.

\subsection{Gauge-invariance}
As discussed above the RPAE-approximation can be shown to produce gauge-invariant results~\cite{lin:77}  for the one-photon processes. This holds when the approximation is used consistently and without truncations. For example, the sum over core orbitals  in Eq.~(\ref{RPA})  cannot  be truncated and the orbital energies should be eigenvalues to the one-particle Hamiltonian used and cannot be replaced with 
experimental ionization energies. With these constraints we are here able to demonstrate gauge invariance also for the two-photon RPAE-approximation as will be seen below.

\section{Results} 
\label{results}
Here we present calculations of atomic delays for photoelectrons emitted along the polarization axis $\mathbf{\hat z}$, as defined in Eqs.~(\ref{phi2})--(\ref{wignerdelay}). 
In Figs~\ref{fig:Nelength}--\ref{fig:final} the horizontal axis  labeled photon energy means the total photon energy absorbed by the photoelectron. In the case of atomic delays this implies the XUV-photon energy plus or minus the laser photon energy for laser absorption or emission, respectively.

\subsection{Neon 2p}
\begin{figure}
\includegraphics[width=0.5\textwidth]{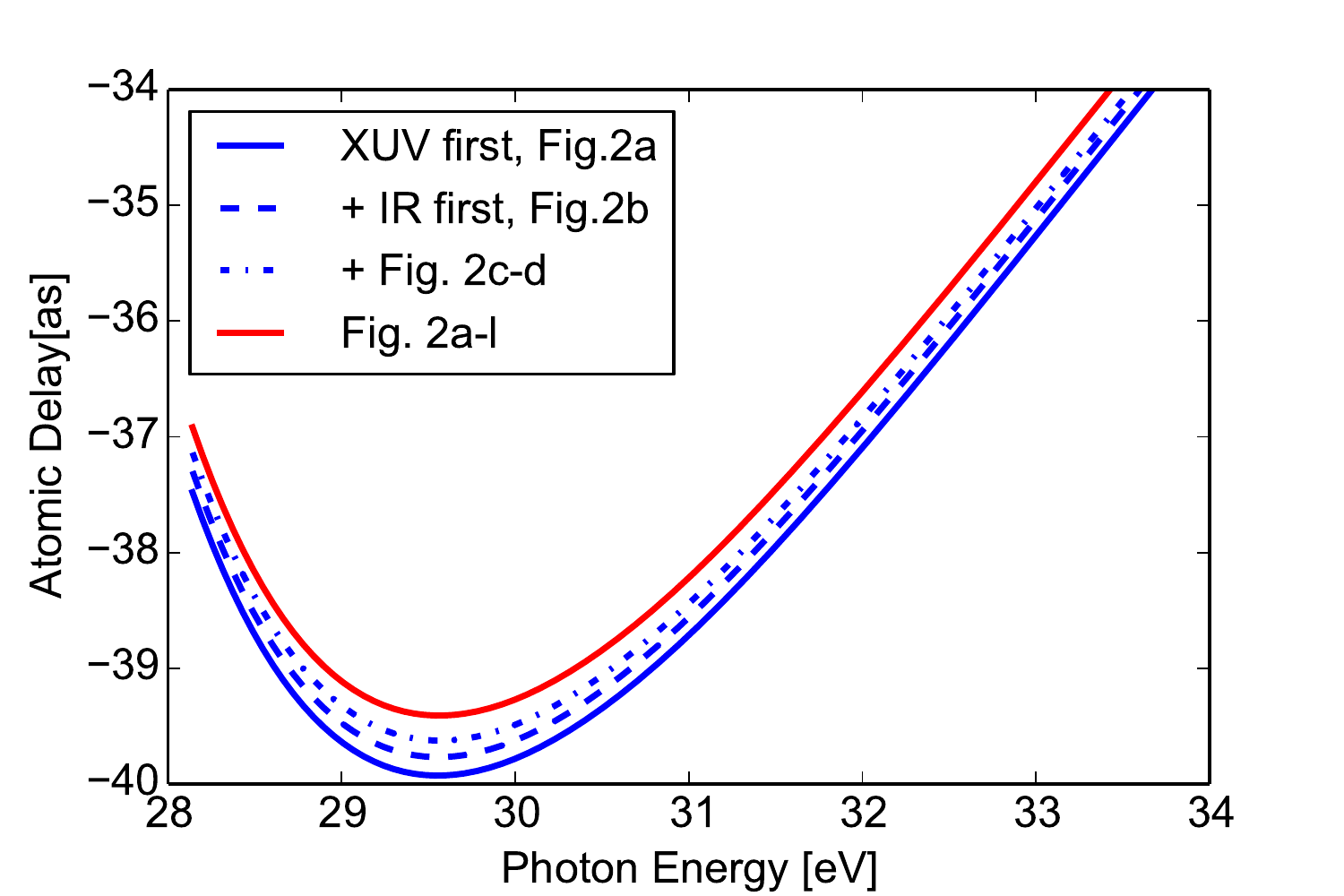}
\caption{
\label{fig:Nelength}
The atomic delay for ionization from Ne $2p$ for electrons emitted along the polarization axis. The delay is   calculated in length gauge. The solid blue line shows the result from Fig.~\ref{fig:M2full}a). The dashed blue line include also Fig.~\ref{fig:M2full}b) and the dashed-dotted blue line also Fig.~\ref{fig:M2full}~c-d). The red sold line shows the final results with the full set of diagrams illustrated in Fig.~\ref{fig:M2full}}. 
\end{figure}

\begin{figure}
\includegraphics[width=0.5\textwidth]{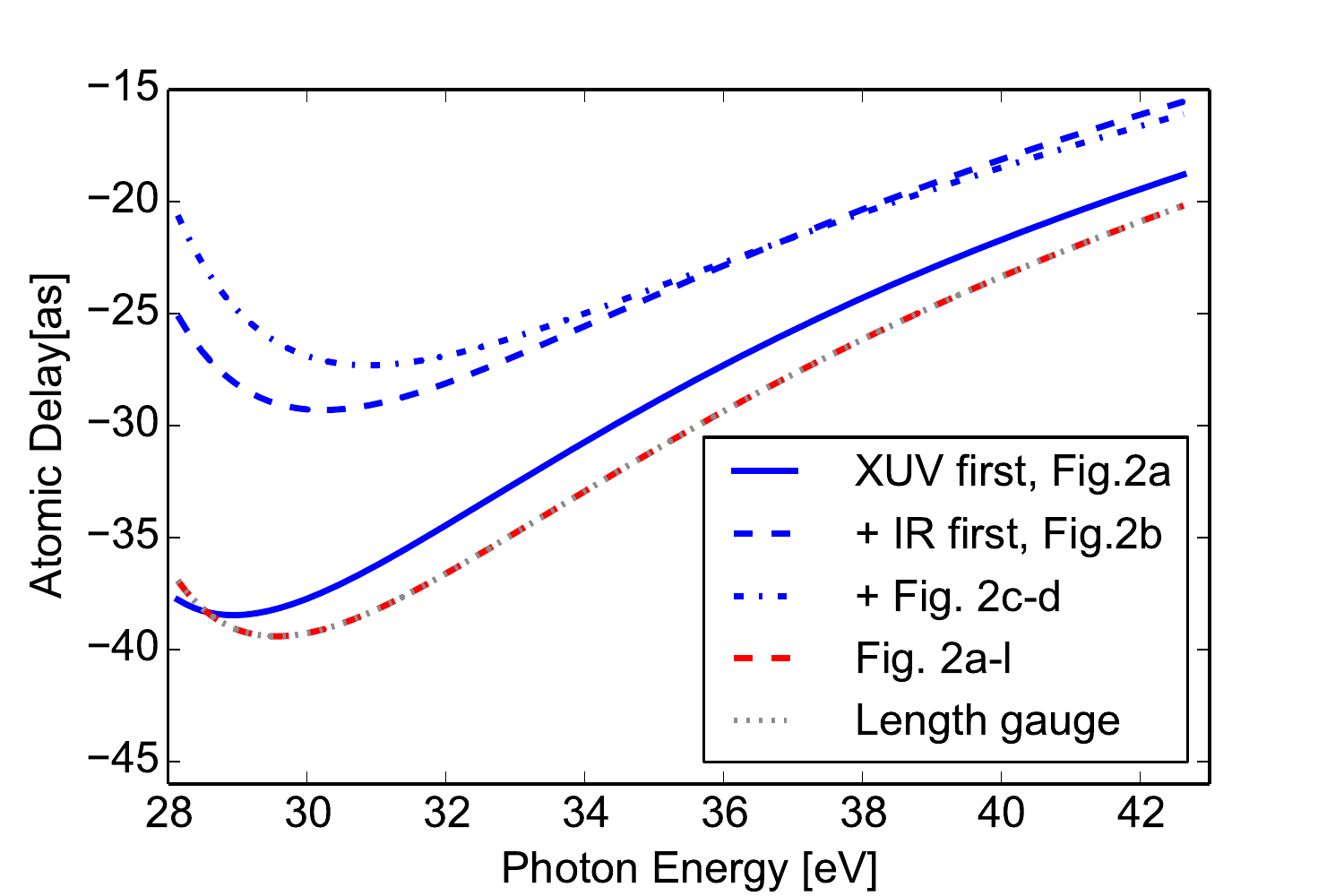}
\caption{
\label{fig:Nevelocity}
The atomic delay for ionization from Ne $2p$ for electrons emitted along the polarization axis. The delay is   calculated in velocity gauge. The solid blue line shows the result from Fig.~\ref{fig:M2full}a). The dashed blue line include also Fig.~\ref{fig:M2full}b) and the dashed-dotted blue line also Fig.~\ref{fig:M2full}c-d). The red dashed line shows the final results with the full set of diagrams illustrated in Fig.~\ref{fig:M2full}. It can be compared to the final result obtained in length gauge (gray dotted line).
}
\end{figure}

The results for photoionization from the $2p$ orbital of neon are presented in Fig.~\ref{fig:Nelength} (length gauge) and Fig.~\ref{fig:Nevelocity} (velocity gauge). The RPAE iterations account for correlation effects from all three orbitals ($1s$, $2s$, $2p$), and the diagrams are evaluated  using HF orbital energies. It is striking that the length gauge result is completely dominated by correlated XUV absorption followed by uncorreleted photoelectron--IR interaction, represented by the diagram in Fig.~\ref{fig:M2full}~(a). Only very small corrections, less than an attosecond, are found from the reversed time-order process (b), uncorrelated hole--field interations (c--d) and general correlated two-photon processes (e--l). This finding is supported by the comparison with experiment in Ref.~\citep{Isinger:2017}, where good agreement was found over a large energy interval in length gauge using only the diagram in Fig.~\ref{fig:M2full}~(a) with experimental values substituted for the HF orbital energies. 
The results are more subtle in velocity gauge. The XUV first with uncorrelated photoelectron--IR interaction appears to be a reasonable approximation that deviates by a few attoseconds from the full calculation, but when the reversed time-order process is added (IR first) the deviation from the full calculation increases. Similarly, adding the hole--field interactions increases the deviation of the atomic delay further. Only the full two-photon RPAE calculation gives identical results in velocity and length gauge as seen in Fig.~\ref{fig:Nevelocity}. The agreement between the gauges can be viewed as a validation test of the implementation.

\subsection{Argon 3p}
The atomic delay for ionization from argon $3p$ is displayed in Fig.~\ref{fig:argonvelocity}.  The RPAE iterations  account for effects from all five orbitals ($1s$, $2s$, $2p$, $3s$, $3p$),
and the diagrams are evaluated 
using HF orbital energies. The delay is larger and changes more dramatically in argon as compared to neon. The velocity gauge result from Fig.~\ref{fig:M2full}~(a) alone underestimate the delay with around 40\,as below the Cooper minimum and overestimate it by more than 50\,as above. Including the full set of diagrams illustrated in Fig.~\ref{fig:M2full}~(a--l)  leads to agreement between the length and velocity results within the numerical accuracy of the calculation. 

In a truncated calculation, where the RPAE iterations account only for effects from the two outer shells $(3s,3p)$, there is a remaining difference between the two gauges, as shown in Fig.~\ref{fig:argon3s3p}. The deviation from the full result is of the same order of magnitude for the two gauges, which implies that there is no clearly preferable gauge for the truncated two-photon RPAE calculation.

\subsection{Argon 3s}
The atomic delay for ionization from argon $3s$ with photoelectrons emitted in the polarization direction is displayed in Fig.~\ref{fig:argon3sExp}. The RPAE iterations  account for effects from all five argon orbitals $(1s,2s,2p,3s,3p)$. While Koopman's theorem states that the binding energy is equal to minus the HF orbital energy, which for $3s$ is $\sim 34.8$~eV, the true ionization energy is only $\sim 29.2$~eV. Therefore, we must substitute the HF orbital energies with experimental values for meaningful comparison with experiments. This simple procedure can be justified, since it corresponds to the inclusion of  additional classes of diagrams \cite{ChengKellyPRA1989}, but only at the price that the results again become gauge dependent. It is  known from one-photon absorption experiments on argon that the cross section of $3s$ is affected by the strong $3p$ photoionization channel through electron correlation effects \cite{Amusia1990}. For two-photon processes a coupling from $3p$ to $3s$ can be directly stimulated by the second photon through the diagrams in Fig.~\ref{fig:M2full}\,(c--d). The question now arises if such hole--field coupling effects can influence the atomic delay in argon?  

\begin{figure}
\includegraphics[width=0.5\textwidth]{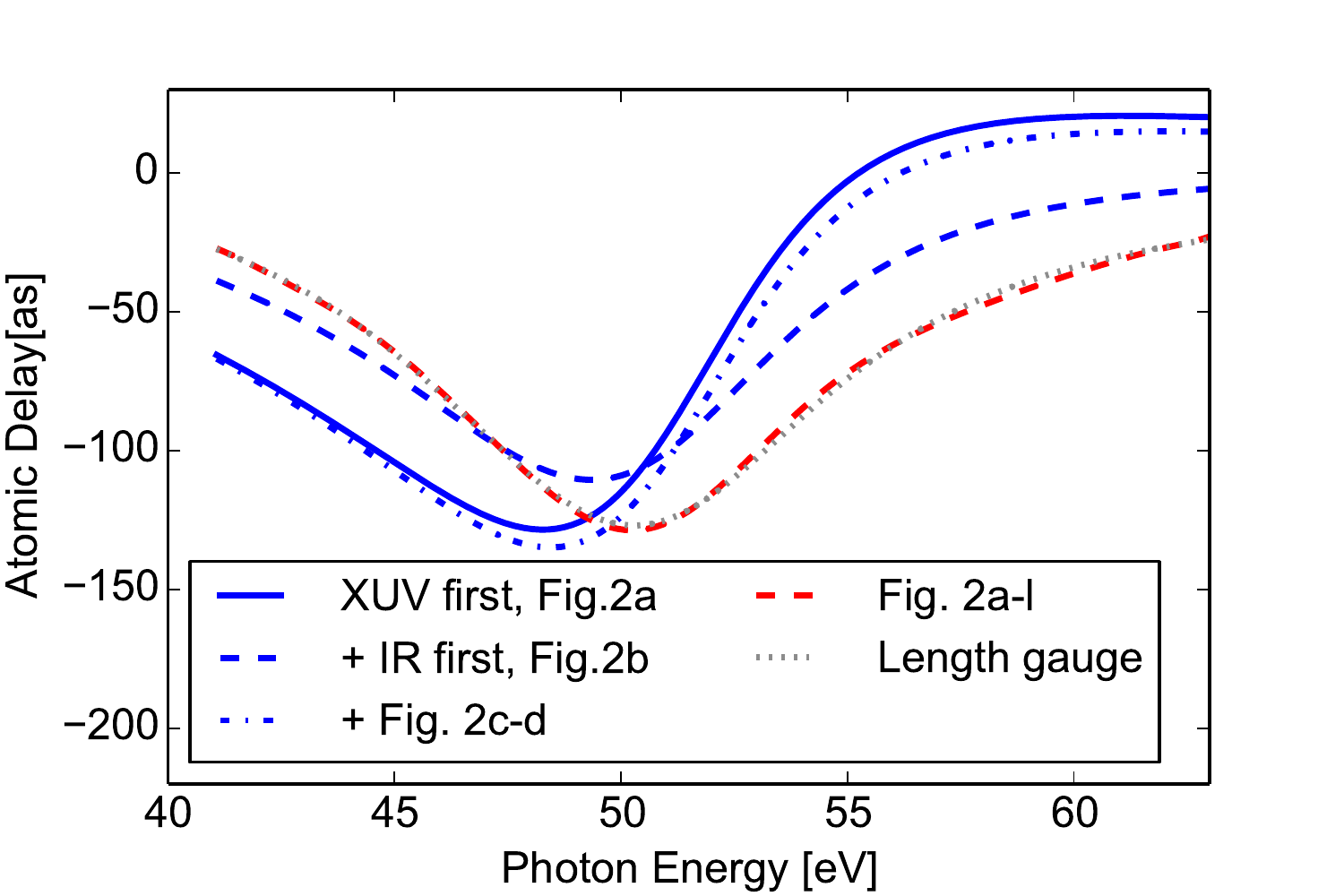}
\caption{
\label{fig:argonvelocity}
The atomic delay for ionization from Ar $3p$ for electrons emitted along the polarization axis.
 The delay is  calculated in velocity gauge. The solid blue line shows the result from Fig.~\ref{fig:M2full}a). The dashed blue line include also Fig.~\ref{fig:M2full}b) and the dashed-dotted blue line also Fig.~\ref{fig:M2full}c-d). The red dashed line shows the final results with the full set of diagrams illustrated in Fig.~\ref{fig:M2full}. It can be compared with the final result obtained in length gauge (gray dotted line).}
\end{figure}

\begin{figure}
\includegraphics[width=0.5\textwidth]{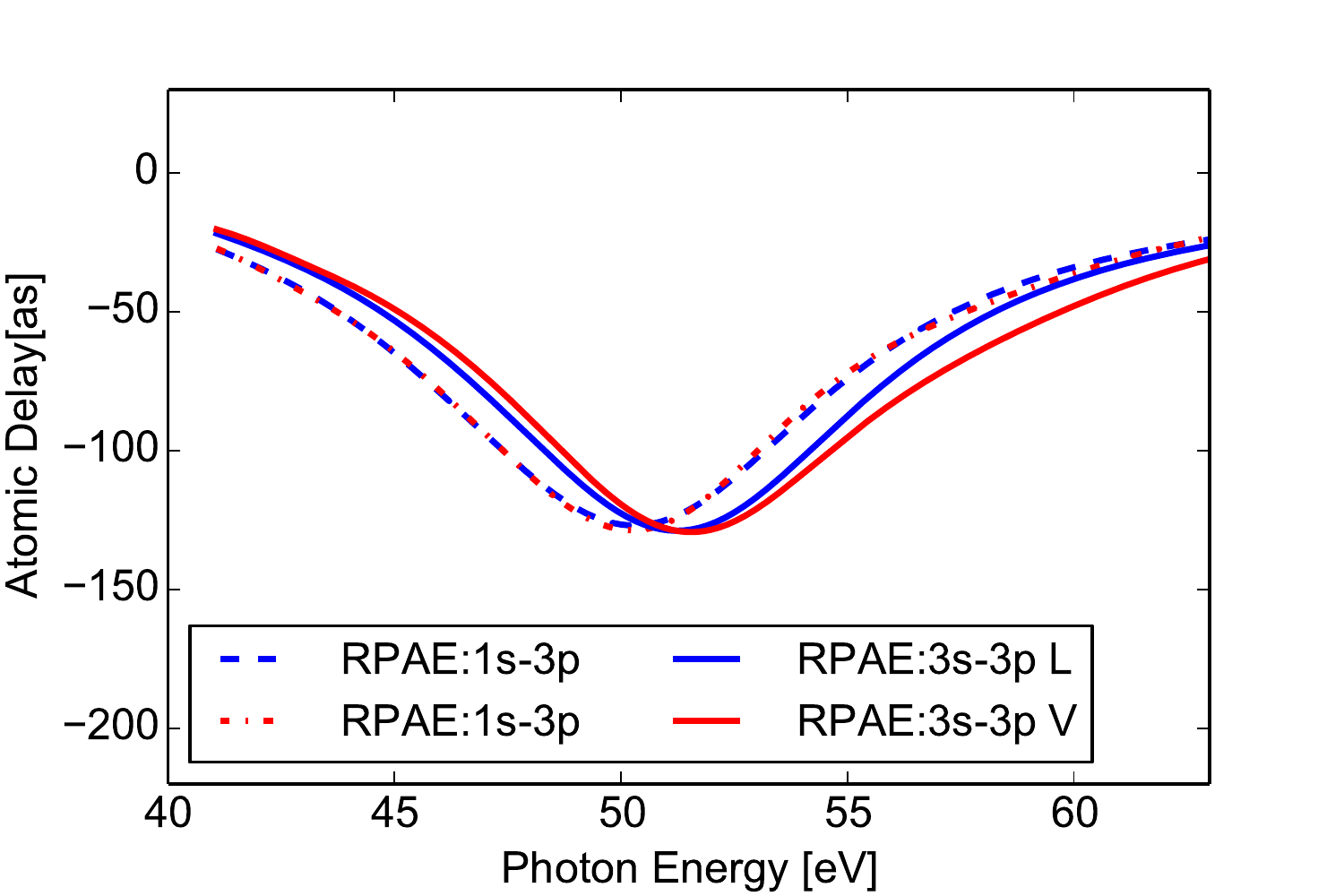}
\caption{
\label{fig:argon3s3p}
The atomic delay for ionization from Ar $3p$ for electrons emitted along the polarization axis using the full set of diagrams illustrated in Fig.~\ref{fig:M2full}. 
The dashed and dashed-dotted lines show the results in both gauges when all five orbitals are included in the two-photon RPAE iterations, while the solid lines show the result with only the two outer orbitals included.}
\end{figure}

\begin{figure}
\includegraphics[width=0.47\textwidth]{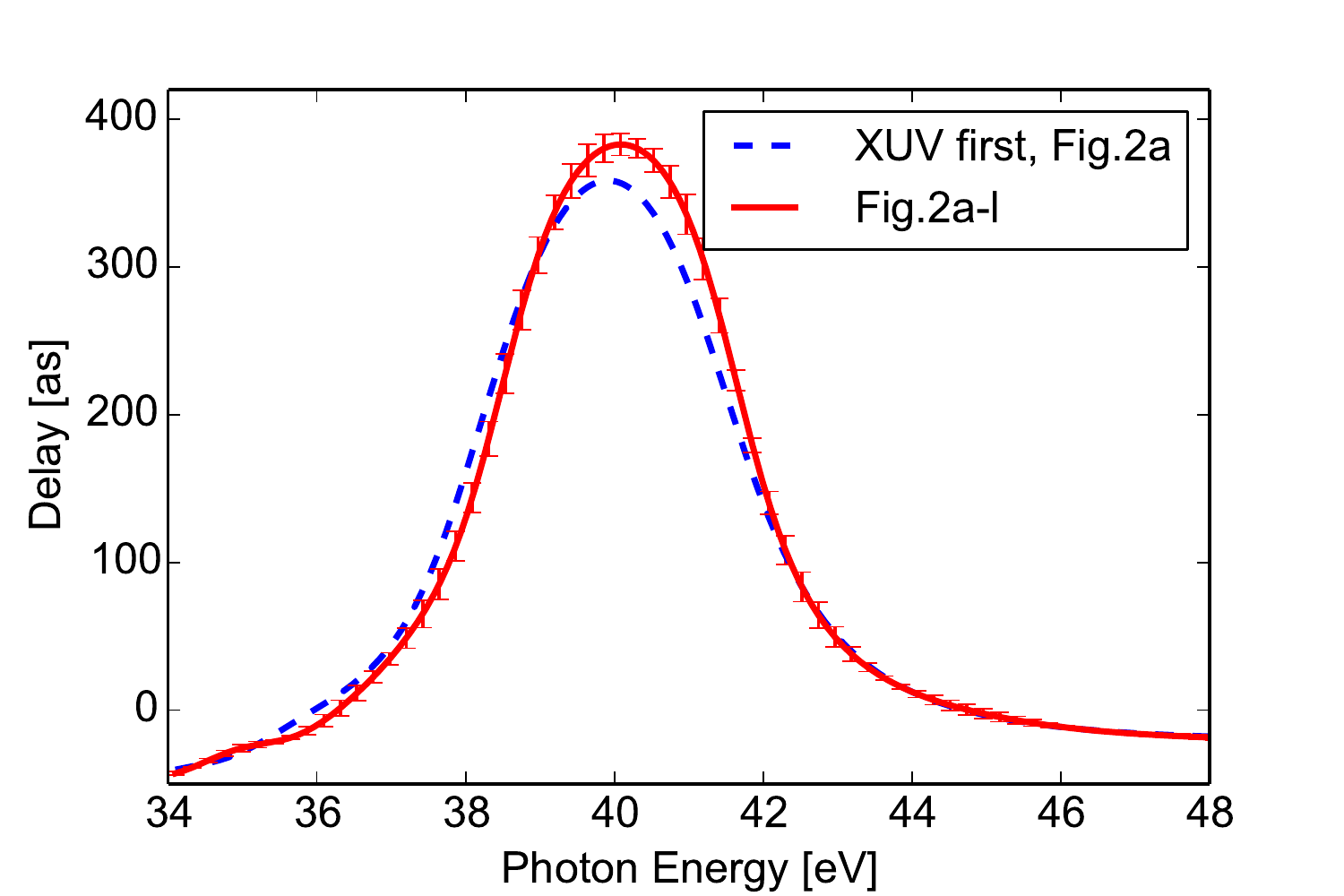}
\caption{The atomic delay for ionization from Ar $3s$ for electrons emitted along the polarization axis. The delay is  calculated in length gauge and the Hartree-Fock orbital energies have been replaced with experimental ionization energies. The dashed blue line shows the result from Fig.~\ref{fig:M2full}a). The solid red line includes  the full set of diagrams illustrated in Fig.~\ref{fig:M2full}. The energy region shown is that of the Cooper minimum in the  $3s$ - cross section.
Very close to this minimum, when the one-photon amplitude goes through zero, the numerical uncertainty grows. The error bars indicate the  spread in the results for different ``break points'' (cf.~\ref{methodC}).
\label{fig:argon3sExp}}
\end{figure}

The Cooper minimum in the $3s$ ionization cross section can be understood as a ``replica'' of the Cooper minimum in the $3p$ ionization channel. In more detail, the $3s$ minimum is caused by an interference effect between the direct path ($3s$) and correlated path ($3p\rightarrow 3s$), which results in very different ionization delays. While the $3p$ delays show a large negative peak (Fig.~\ref{fig:argonvelocity}), the $3s$ delays show a large positive peak shown in  Fig.~\ref{fig:argon3sExp}. This conclusion is consistent with prior works based on RPAE \citep{DahlstromPRA2012,KheifetsPRA2013,DahlstromJPB2014,PhysRevA.97.063404} and Time-Dependent Local-Density-Approximation (TDLDA) ~\cite{PhysRevA.91.063415}.  
Oddly, the large positive peak has not been observed in experiments ~\citep{KlunderPRL2011,GuenotPRA2012}, while the negative peak has been reproduced experimentally using RABBIT \cite{SchounPRL2014,PalatchiJPB2014}. 
In our earlier studies of atomic delays, we have only accounted for Fig.~\ref{fig:M2full}~(a) and we have found that both the sign and position of the $3s$ delay peak is sensitive to correlation effects \cite{DahlstromJPB2014}. Here, we find that the contributions from the remaining diagrams in Fig.~\ref{fig:M2full}, are not insignificant for $3s$ in argon, as seen in Fig.~\ref{fig:argon3sExp}, and that the main additional contributions come from the hole--field coupling in Fig.~\ref{fig:M2full}~(c). However, the sum of all diagrams in our complete two-photon RPAE calculation does not resolve the discrepancy with argon $3s$ experiments at the Cooper minimum because the sign of our final delay peak remains positive and its position is not significantly altered (much less than an electron volt).  

Very close to the Cooper minimum, where the one-photon matrix element goes through zero, it has been found harder to achieve good numerical accuracy. The scatter of break-points, see Sec.~\ref{methodC}, is indicated by error-bars in Fig.~\ref{fig:argon3sExp}.

\subsection{Argon $3s-3p$}
\begin{figure}
\includegraphics[width=0.47\textwidth]{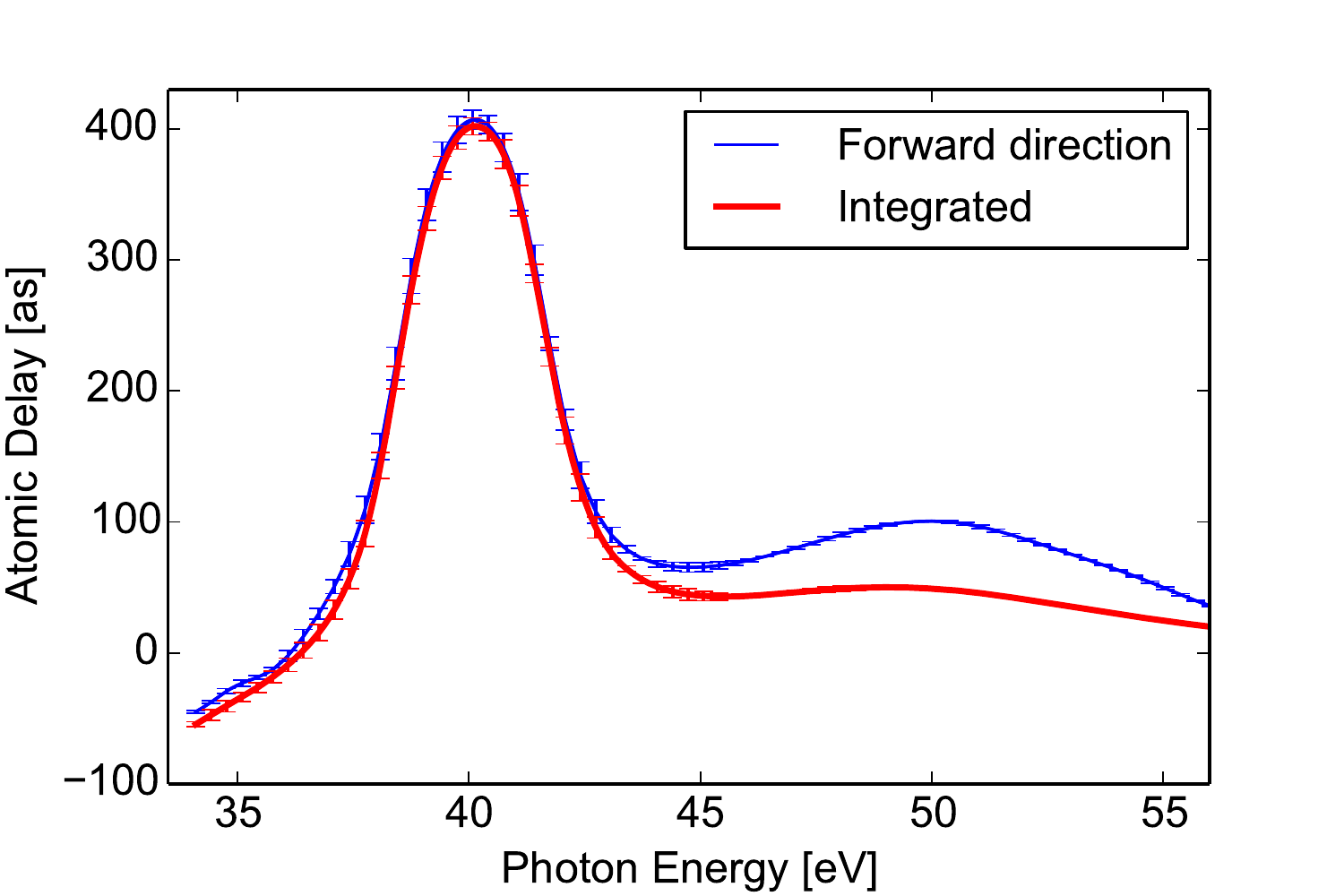}
\caption{ The measurable difference between the atomic delays for 
Ar $3s$ and $3p$. The blue line shows the delay-difference for electrons emitted along the polarization axis of the laser field and the red line the angle-integrated result. The error bars indicate the  spread in the results for different ``break points'' (cf.~\ref{methodC}).
\label{fig:final}}
\end{figure}

Finally, we show the difference in atomic delay for photoelectrons ionized from the two outer orbitals in argon, $\tau_A^{(3s)}-\tau_A^{(3p)}$, in  Fig.~\ref{fig:final}. Here we display both the calculated result for electrons emitted in the polarization direction $\mathbf{\hat z}$,  and for angular integrated detection, which is the configuration used in current RABBIT  experiments~\cite{KlunderPRL2011,GuenotPRA2012,salieres:priv}. We find that the atomic delay difference is not affected by the choice of detection in the region of the $3s$ Cooper minimum at $\sim 40$\,eV. In contrast, the atomic delay difference is strongly altered close to the $3p$ Cooper minimum at $\sim 50$\,eV due to the choice of detection, and the delay peak is reduced due to angular integration in agreement with the experimental angle-integrated results for argon $3p$ \cite{PalatchiJPB2014} and calculations  \cite{DahlstromJPB2014corr}. 

\section{Discussion}
\label{sec:discussion}
\subsection{Gauge dependence}
In general, our calculations show that much larger contributions  arise from the reversed time-order, where the less energetic
photon is absorbed first, in velocity gauge than in length gauge. This can be understood using a simple analytical calculation. For a  Hamiltonian, $h$, with  a local potential the length and velocity form of the dipole operator satisfy
\begin{align}
\left[h, e\mathbf{r}\right]=\frac{-i e \hbar}{m}\mathbf{p}.
\end{align}
By assuming the dipole approximation, the vector potential for a given angular frequency and mode can be written as:
\begin{align}
A(t) = 2iA_\Omega\sin\Omega t = 
A_\Omega \left( e^{i\Omega t} - e^{-i\Omega t}\right), 
\end{align}
where the two complex exponents can be physically interpreted as the drivers for emission and absorption of laser photon by the atom, respectively.  
Using the relation, $\mathbf{E}=-d\mathbf{A}/dt$, the expression for the electric field amplitude along the polarization axis $\mathbf{\hat z}$ is 
\begin{align}
E_{\Omega}  = \mp i \Omega A_\Omega,
\end{align}
for emission and absorption, respectively. 
The gauge invariance of on-shell matrix elements follows from
\begin{align}
\frac{e \hbar}{m} \langle s \mid p_z \mid r \rangle A_\Omega = \frac{e}{\mp \Omega} \left( \epsilon_s -\epsilon_r \right) \langle s \mid   z \mid r \rangle E_\Omega,
\label{offdiagonal}
\end{align}
provided that  
$\left( \epsilon_s -\epsilon_r \right) = -\Omega$ for emission
(photon creation)  and
$\left( \epsilon_s -\epsilon_r \right) =\Omega$ for absorption (photon annihilation). 

Off-shell matrix elements are generally different. 
Consider the two-photon transition matrix element from initial state $\mid 0 \rangle$ to a final state with $\epsilon_f = \epsilon_0 + \Omega_1 + \Omega_2$   via an intermediate state $\mid i \rangle$: 
\begin{align}
\frac{
\langle f \mid d_{\Omega_2} \mid i \rangle \langle i  \mid d_{\Omega_1} \mid 0\rangle}
{\epsilon_0 + \Omega_1 - \epsilon_i}
\end{align}
the velocity gauge result is then a factor
\begin{align}
\label{velfactor}
\left(  \epsilon_i -  \epsilon_0\right)
\left(  \epsilon_0 + \Omega_1 + \Omega_2 -  \epsilon_i \right)/\Omega_1 \Omega_2
\end{align}
times the length gauge result.   Eq.~(\ref{velfactor})   has a maximum at $\epsilon_i = \epsilon_0 +  \left(\Omega_1 + \Omega_2\right)/2 $, and 
at this maximum it amounts to
\begin{align}
\left(\frac{\Omega_1 + \Omega_2}{2}\right)^2
\frac{1}{\Omega_1 \Omega_2}\approx \Omega_>/4 \Omega_< .
\end{align}
Therefore, we expect to find large differences for individual diagrams when, as in in typical RABBIT situation, the XUV photon has an energy of $20-40$ IR photons. We have indeed seen that the second time-order (TO2), which is always off-shell, is much more important in velocity gauge than in length gauge. Contributions for intermediate excited states  and continuum states closely above the ionization threshold are likely to dominate, and for those the enhancement factor in Eq.~(\ref{velfactor}) will be of quite some importance. 

\subsection{Universality of cc-delays in argon}
\begin{figure}
\includegraphics[width=0.47\textwidth]{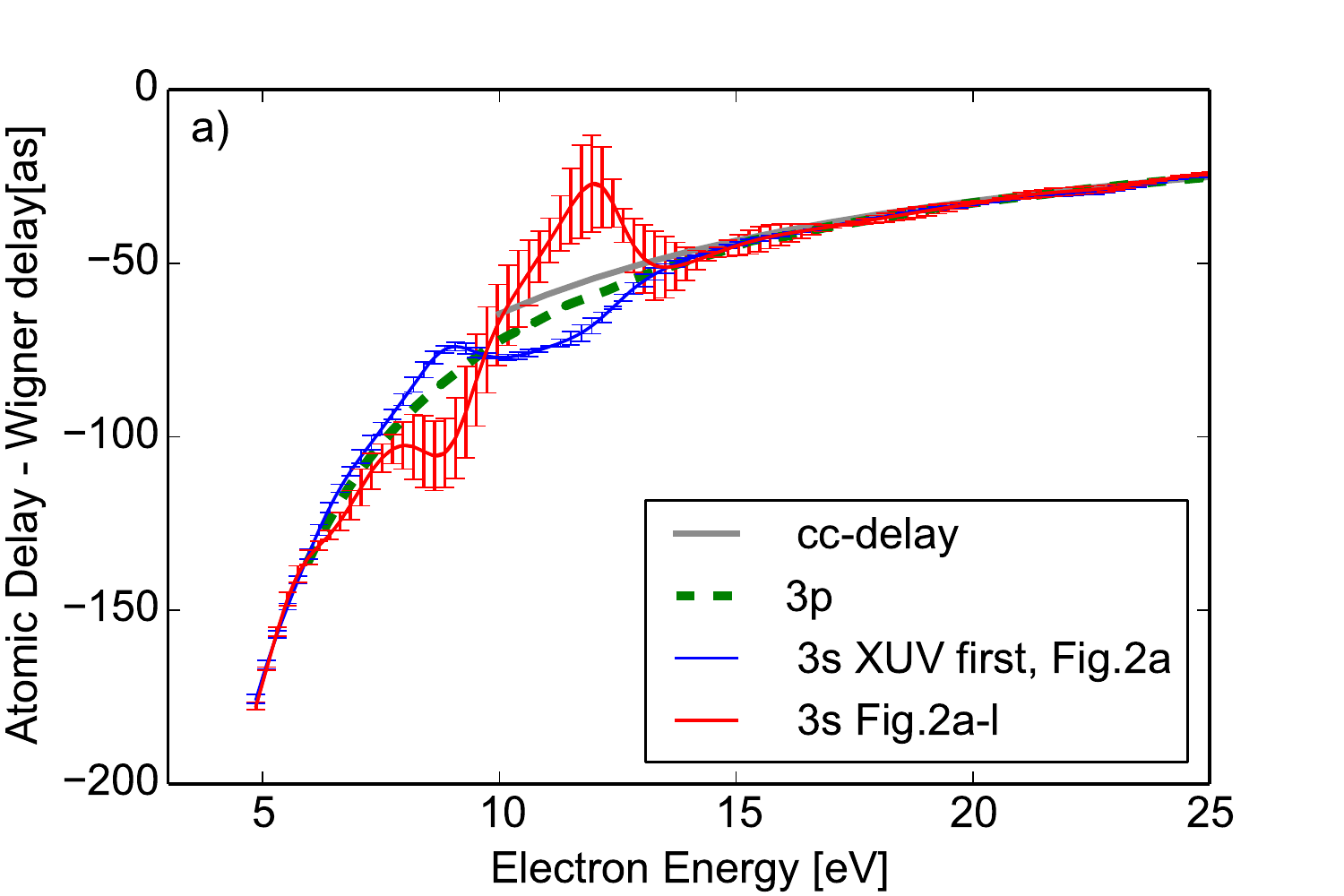}
\includegraphics[width=0.47\textwidth]{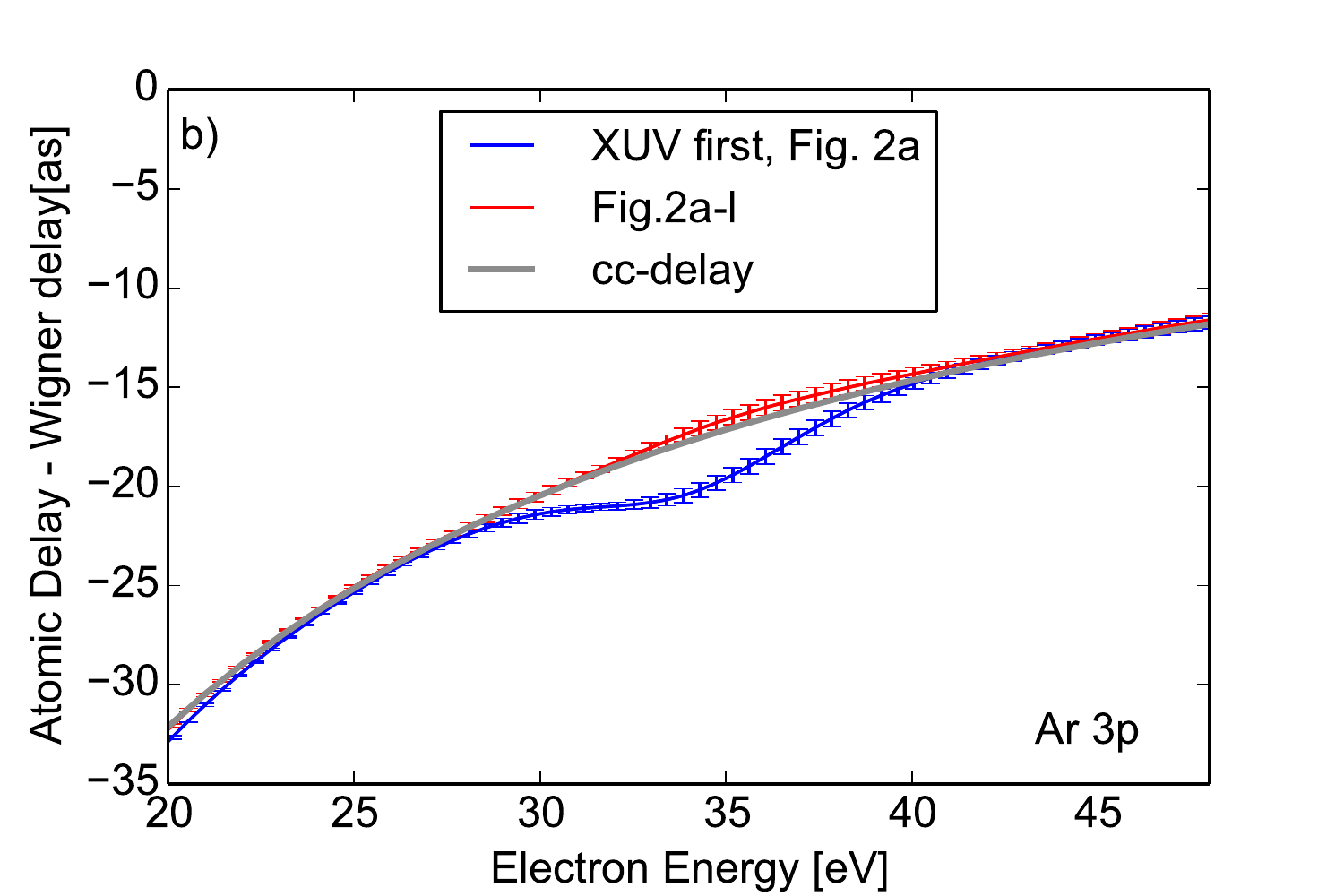}
\caption{ The difference between the atomic delay and the Wigner delay for argon calculated in length gauge. (a) The energy region around the Ar~$3s$ Cooper-minimum. The solid gray line shows the continuum-continuum delay~\cite{DahlstromJPB2012}. The results for Ar~$3p$ (dashed green line) agrees well with the continuum-continuum delay in higher energy range. For Ar~$3s$ there are clear deviations from the universal curve. Close to the $3s$ Cooper-minimum, where the one-photon amplitude goes through zero, the numerical uncertainty grows, especially for the full two-photon RPAE results.  The error bars reflects the scatter between the different ``break points''. (b) The energy region of the $3p$ Cooper-minimum. Here there is a small deviation from the continuum-continuum delay for Ar~$3p$ in the simplest approximation (Fig~2a), while the result with full two-photon RPAE reproduces the universal curve nicely. The error bars indicate the  spread in the results for different ``break points'' (cf.~\ref{methodC}).
\label{fig:taucc}}
\end{figure}

In Fig.~\ref{fig:taucc} we show the {\it difference} between atomic delay and  Wigner delay, $\tau_A-\tau_W$, for argon from orbital  $3p$ and $3s$. Panel (a) shows the low energy region with the $3s$ Cooper minimum, while panel (b) shows the high energy region with the $3p$ Cooper mininum. The approximate continuum--continuum delay, $\tau_{CC}$, is shown for comparison and it is calculated using the analytical expression of Eq.~(100) from Ref.~\cite{DahlstromJPB2012}. The analytical cc-delay takes into account both long-range phase effects and long-range amplitude effects based on the Wentzel-Kramers-Brillouin (WKB) approximation, which gradually breaks down at low kinetic energies \cite{DahlstromJPB2012}. Recently, excellent numerical agreement between $\tau_A-\tau_W$ for argon $3p$ and neon $2p$ was reported at very low kinetic energies using the diagram in Fig.~\ref{fig:M2full}~(a) \cite{LindrothPRA2017}. This  suggested that the concept of ``universality'' goes beyond the analytical predictions of Ref.~\cite{DahlstromJPB2012}, down to much lower energies close to the ionization threshold, where the WKB approach is not applicable.  

Surprisingly, the delay difference for argon $3s$ does {\it not} follow the universal curve [indicated by green dashed curve with data for $3p$ in Fig.~\ref{fig:taucc}~(a)], but instead shows irregular deviations at low electron energies close to the $3s$ Cooper minimum in Fig.~\ref{fig:taucc}~(a). The one-photon amplitude goes through zero at a photon energy of $\sim 40$\,eV, which results in increased numerical uncertainty. The errorbars in Fig.~\ref{fig:taucc} reflect the scatter between the different ``break points'', c.f.~Sec.~\ref{methodC}, and they signify that the observed deviations of $3s$ from $3p$ are real and that the universal trend is indeed broken, despite our limited numerical accuracy in this region.  When only laser-stimulated continuum transitions are included in the calculation [Fig.~\ref{fig:M2full}~(a)], the deviation from the universal curve is not very large. The major part of the deviation comes from the remaining diagrams [Fig.~\ref{fig:M2full}~(b)--(l)], which suggests the importance of additional ways for the atom to interact with the fields when the single-photon XUV ionization process goes to zero.
Because there are no resonances in the energy region shown for argon with RPAE, the irregular behaviour must be associated to the argon $3s$ Cooper minimum. Above {\it and} below the $3s$ Cooper minima, we find that the $3s$ delay difference agrees with the universal curve of $3p$, which indicates that correlation effects beyond the diagram in Fig.~\ref{fig:M2full}~(a), are significant only close to the exact photon energy region where the otherwise dominant one-photon correlated XUV ionization process vanishes. 

What about the $3p$ Cooper minimum? This minimum is a ``typical'' Cooper minimum \cite{cooper:1962} that arises due to a zero in the dipole transition $p \rightarrow d$ in XUV photoionization. The partial $3p$ cross-section does, however, not go to zero because the $p \rightarrow s$ dipole transition remains finite at all XUV energies. The deviation of $3p$ from the universal curve [indicated by the gray full curve in Fig.~\ref{fig:taucc}~(b)] is found to be small when laser stimulated continuum transitions are considered [Fig.~\ref{fig:M2full}~(a)] and very surprisingly even smaller when the full set of diagrams are included [Fig.~\ref{fig:M2full}~(a)--(l)].  

Similar small deviations from the universal curve can be spotted in Fig.~5 of Ref.~\cite{DahlstromPRA2012} for both argon $3s$ and $3p$, but because the effects are small compared to the associated atomic delays, they were not given much attention.    
More recently, irregular deviations from the ``universal'' curve of up to 20\,as close to the argon $3p$ Cooper minimum was reported in Ref.~\cite{BrayPRA2018}. Our calculations show that such deviations are orders of magnitude too large and that they most likely arise due to an inconsistent description of combined correlation and field effects. The reduction of $3p$ deviation from the universal curve down to sub-attosecond precision in Fig.~\ref{fig:final}~(b) is most likely due to our improved description of the final state, where the effective spherical projected potential is substituted by self-consistent final state correlation effects  [Fig.~\ref{fig:M2full}~(k)].       

\section{Conclusion}
\label{sec:conclusion}
In this work we have shown that full two-photon RPAE calculations of atomic delays give gauge invariant results and that effects beyond the universal IR--photoelectron continuum--continuum transitions are rare, but do occur in special cases. In particular, we have found that the argon $3s$ Cooper minimum suffers from a non-universal delay because the correlated XUV dipole moment for photoionization vanishes, so that  other processes, including XUV--hole interaction, may play an important role for the two-photon process. In contrast, we find that there are no such deviations from the universal delay curve for $3p$ in argon. Any $3p$ deviations that we find are on a sub-attosecond time scale, which disproves the strong $3p$ deviations recently proposed in Ref.~\cite{BrayPRA2018} using a hybrid RPAE+TDSE approach.  

Despite our best efforts, we have not been able to explain the discrepancy between theory and experiment for the argon $3s-3p$ atomic delays. This is because the full two-photon RPAE calculation still shows a positive peak in the atomic delay peak that is absent in experiments \cite{KlunderPRL2011,GuenotPRA2012,salieres:priv}. We note that recent simulations using Time-Dependent Density Functional Theory (TDDFT)~\citep{Sato2018} have generated results for the argon $3s-3p$ delay, in better agreement with experiments in this energy region. The authors of Ref.~\citep{Sato2018} attribute this success  to their consistent treatment of the interaction with both light fields, as compared to the hybrid TDLDA+CC result in Ref.~\cite{PhysRevA.91.063415}. However, the results of hybrid approaches, such as RPAE+CC \cite{KheifetsPRA2013} and TDLDA+CC \cite{PhysRevA.91.063415} where the effect of the laser field is treated by a simple time shift given by analytical formulas  \cite{DahlstromJPB2012,PazourekFD2013}, are mostly consistent with our new results. We cannot support the conclusion that an inconsistent description of the fields is the reason of the disagreement with experiments, because our present study  does imply a consistent treatment of many-body effects for both fields. Still, it is hard to compare the many-body effects included with TDDFT (or TDLDA) with the present calculation and it is very well possible that the difference between the calculations lies here. 

In closing, we wish to stress that XUV photoionization of argon is associated with strong satellite peaks that have not been considered in the present work, but have been studied in detail by Wijesundera and Kelly using many-body perturbation theory for the one-photon ionization process \cite{kellyPRA1987}. A direct comparison between the partial cross-section for  $3s^{-1}$ and the dominant satellite $3p^{-2}(^1D)3d$ from Ref.~\cite{kellyPRA1987}, shows that the satellite process does dominate in the photoelectron energy region of the $3s$ Cooper minimum with a cross-section of $\sim 0.02$\,Mb as compared to our present value for the $3s$ partial cross-section of $\sim0.003$\,Mb located at a photon energy of $\sim 40$\,eV using RPAE with all atomic orbitals and experimental energies in length gauge. Therefore, in order to better understand the discrepancy between experiments and theory it would be helpful to acquire atomic delays for larger energy ranges, but also to study the one-photon and two-photon partial cross-sections for $3s$ to be able to locate the exact position of the associated Cooper minima. Continued studies of shake-up processes in attosecond science, that go beyond the hybrid MCHF+CC approach of Ref.~\cite{FeistPRA2014}, is desirable and maybe the right path to solve the long-standing argon delay puzzle.      

\section*{Acknowledgments}
The authors acknowledge support from the Knut and Alice Wallenberg Foundation and the Swedish Research
Council, Grant No. 2014-3724, 2016-03789 and 2018-03845.
%

\end{document}